\documentclass[journal]{IEEEtran}

\usepackage{epsfig, latexsym, amssymb, amsmath, url}
\usepackage{graphics}
\usepackage{algorithm}
\usepackage{algorithmic}
\usepackage{subfigure}

\newtheorem{prop}{Proposition}[section]
\renewcommand{\theprop}{\arabic{section}.\arabic{prop}}
\newtheorem{cor}{Corollary}
\renewcommand{\thecor}{\arabic{section}.\arabic{cor}}
\newtheorem{lm}{Lemma}

\newtheorem{thm}{Theorem}

\newcommand{\bthm}{\begin{thm}}
\newcommand{\ethm}{\end{thm}}

\newcommand{\bcor}{\begin{cor}}
\newcommand{\ecor}{\end{cor}}
\newcommand{\bprop}{\begin{prop}}
\newcommand{\eprop}{\end{prop}}
\newcommand{\blm}{\begin{lm}}
\newcommand{\elm}{\end{lm}}
\newcommand{\beq}{\begin{equation}}
\newcommand{\eeq}{\end{equation}}
\newcommand{\ber}{\begin{eqnarray}}
\newcommand{\eer}{\end{eqnarray}}

\newenvironment{proof1}{\begin{trivlist}\item[]{\bf Proof:\hspace{2mm}}}{\hfill$\blackbox$\end{trivlist}}

\newcommand{\blackbox}{\vrule height7pt width5pt depth1pt}

\newcommand{\bit}{\begin{itemize}}
\newcommand{\eit}{\end{itemize}}
\newcommand{\ben}{\begin{enumerate}}
\newcommand{\een}{\end{enumerate}}
\newcommand{\bdesc}{\begin{description}}
\newcommand{\edesc}{\end{description}}
\newcommand{\beqarrn}{\begin{eqnarray*}}
\newcommand{\eeqarrn}{\end{eqnarray*}}
\newenvironment{proofof}[1]{\begin{trivlist}\item[]{\bf Proof of #1:\hspace{2mm}
}}{\hfill\blackbox\end{trivlist}}
\newcommand{\bproofof}{\begin{proofof}}
\newcommand{\eproofof}{\end{proofof}}
\newenvironment{rem}{\begin{trivlist}\item[]{\bf
Remark:}\hspace{4mm}}{\end{trivlist}}
\newcommand{\brem}{\begin{rem}}
\newcommand{\erem}{\end{rem}}
\newenvironment{rems}{\begin{trivlist}\item[]{\bf
Remarks}\begin{itemize}}{\end{itemize}\end{trivlist}}
\newcommand{\brems}{\begin{rems}}
\newcommand{\erems}{\end{rems}}
\newtheorem{fact}{Fact}
\newcommand{\bfact}{\begin{fact}}
\newcommand{\efact}{\end{fact}}
\newtheorem{examp}{Example}
\newcommand{\bexamp}{\begin{examp}\rm}
\newcommand{\eexamp}{\end{examp}}
\newtheorem{defn}{Definition}
\newcommand{\bdefn}{\begin{defn}\rm}
\newcommand{\edefn}{\end{defn}}

\newtheorem{prob}{Problem}
\newcommand{\bprob}{\begin{prob}}
\newcommand{\eprob}{\end{prob}}

\newcommand{\bvtm}{\begin{verbatim}}
\newcommand{\bfig}{\begin{figure}}
\newcommand{\efig}{\end{figure}}
\newcommand{\bcen}{\begin{center}}
\newcommand{\ecen}{\end{center}}

\long\def\comment#1{}

\def \n2{{N_0 \over 2}}

\newcommand{\bP}[1]{{\mathbb{P}}\left[{#1}\right]}
\newcommand{\bE}[1]{{\mathbb{E}}\left[{#1}\right]}

\newcommand{\1}[1]{{\bf 1}\left[#1\right]}

\def \h5{\hspace{0.5in}}

\begin{document}

\title{Conjoining Speeds up Information Diffusion in Overlaying Social-Physical Networks}


\author{
\IEEEauthorblockN{Osman Ya\u{g}an\IEEEauthorrefmark{1}, Dajun Qian\IEEEauthorrefmark{2},
  Junshan Zhang\IEEEauthorrefmark{2}, and Douglas Cochran\IEEEauthorrefmark{2}}\\
 \IEEEauthorblockA{\IEEEauthorrefmark{1}
CyLab, Carnegie Mellon University, Pittsburgh, PA 15213. E-mail: {\tt{oyagan@andrew.cmu.edu}}}
\\
 \IEEEauthorblockA{\IEEEauthorrefmark{2}
School of ECEE,
Arizona State University, Tempe, AZ  85287}
\\
 {\tt  \{dqian, junshan.zhang, cochran\}}{\tt @asu.edu}
\thanks{This research was supported in part by the U.S.~National Science
Foundation under NSF grants No. CNS-0905603, CNS-0917087, and by the U.S.~Defense Threat Reduction Agency under DTRA
grant HDTRA1-09-1-0032. The material in this paper was presented in part
at the 46th Annual Conference on Information
Sciences and
Systems, Princeton (NJ), March 2012.  }
}

\maketitle

\begin{abstract}
We study the diffusion of information in an overlaying
social-physical network. Specifically, we consider the following set-up:
There is a physical
information network where information spreads amongst people
through {\em conventional} communication media (e.g.,
face-to-face communication, phone calls), {\em and} conjoint to
this physical network, there are online social
networks where information spreads via web sites such 
as Facebook, Twitter, FriendFeed, YouTube, etc. We quantify the size and the
critical threshold of information epidemics in this conjoint
social-physical network by assuming that information diffuses
according to the SIR epidemic model. One interesting finding is
that even if there is no percolation in the individual networks,
percolation (i.e., information epidemics) can take place in the
conjoint social-physical network. We also show, both analytically
and experimentally, that the fraction of individuals who receive
an item of information (started from an arbitrary node) is
significantly larger in the conjoint social-physical network case,
as compared to the case where the networks are disjoint. These
findings reveal that conjoining the physical network with online
social networks can have a dramatic impact on the speed and scale
of information diffusion.

\end{abstract}


\section{Introduction}

\subsection{Motivation}

Modern society relies on basic physical network infrastructures,
such as power stations, telecommunication networks and
transportation systems. Recently, due to advances in communication
technologies and cyber-physical systems, these infrastructures
have become increasingly dependent on one another and have emerged
as {\em interdependent networks} \cite{CPS}. One archetypal
example of such coupled systems is the smart grid where the power
stations and the communication network controlling them are
coupled together. See the pioneering work of Buldyrev et al.
\cite{Buldyrev} as well as \cite{ChoGohKim,
CohenHavlin,
Vespignani,YaganQianZhangCochranLong}
for a diverse set of models on coupled networks.

Apart from physical infrastructure networks, coupling can also be
observed between different types of social networks.
Traditionally, people are tied together in a {\em physical}
information network through old-fashioned communication media,
such as face-to-face interactions. On the other hand, recent
advances of Internet and mobile communication technologies have
enabled people to be connected more closely through {\em online}
social networks. Indeed, people can now interact through e-mail or
online chatting, or communicate through a Web $2.0$ website such
as Facebook, Twitter, FriendFeed, YouTube, etc. Clearly, the
physical information network and online social networks are {\em
not} completely separate since people may participate in two or
more of these
networks at the same time. 
For instance, a person can forward a message to his/her online
friends via Facebook and Twitter upon receiving it from someone through
face-to-face communication. As a result, the information spread in
one  network may trigger the propagation in another network, and
may result in a possible cascade of information. One conjecture is
that due to this {\em coupling} between the physical and online
social networks, today's breaking news (and information in
general) can spread at an unprecedented speed throughout the
population, and this is the main subject of the current study.

Information cascades over coupled networks can deeply
influence the patterns of social behavior. Indeed, people have become
increasingly aware of the fundamental role of the coupled
social-physical network\footnote{Throughout, we sometimes refer to
the physical information network simply as the physical network,
whereas we refer to online social networks simply as social
networks. Hence the term coupled (or overlaying, or conjoint)
social-physical network.} as a medium for the spread of not only
information, but also ideas and influence. Twitter has emerged as
an ultra-fast source of news \cite{WebSite} and Facebook has
attracted major businesses and politicians for advertising
products or candidates. Several music groups or singers have
gained international fame by uploading videos to YouTube. In
almost all cases, a new video uploaded to YouTube, a rumor started
in Facebook or Twitter, or a political movement advertised through
online social networks, either dies out quickly or reaches a
significant proportion of the population. In order to fully
understand the extent to which these events happen, it is of great
interest to consider the combined behavior of the physical
information network and online social networks.

\subsection{Related Work}

Despite the fact that information diffusion has received a great deal of research interest
from various disciplines for over a decade, there has been little study on
the analysis of information diffusion across {\em coupled} networks;
most of the works consider information propagation only within a single 
network. The existing literature on this topic 
is much too broad to survey here, but we will attempt
to cover the works that are most relevant to our study. 
To this end, existing studies
can be {\em roughly} classified into two categories. The first type of studies
\cite{LermanGhosh,KimYoneki,LinBarabasi,
BakshyRosennMarlowAdamic,Liben-NowellKleinberg,LeskovecMcGlohonFaloutsos}
are {\em empirical} and analyze various aspects of information diffusion using large-scale
datasets from existing online social networks. Some of the interesting questions that
have been raised (and answered) 
in these references include
\lq\lq What are the roles of behavioral properties of the individuals and the strength of their ties
in the dynamics of information diffusion" \cite{LinBarabasi,BakshyRosennMarlowAdamic},
\lq\lq 
How do blogs influence each other?" \cite{LeskovecMcGlohonFaloutsos}, and
 \lq\lq How does the topology of the underlying social 
network effect the spread of information?"
 \cite{BakshyRosennMarlowAdamic}.

The second type of studies \cite{AndersonMay,Bailey,Hethcote,KupermanAbramson,MooreNewman,
Pastor-SatorrasVespignani,newman2002spread}
build mathematical models to analyze the mechanisms by which 
information diffuses across the population.
These references study the spread of {\em diseases} (rather than information) 
 in small-world networks
\cite{KupermanAbramson,MooreNewman}, scale-free networks
\cite{Pastor-SatorrasVespignani}, and networks with arbitrary
degree distributions \cite{newman2002spread}. However,
by the well-known analogy between the spread of diseases and information
\cite{Blogspace,Worms,P2P}, their results also apply in the context of information
diffusion. Another notable work in this group is \cite{OstilliYonekiLeungMendes}
which studies the spread of rumors in a network with multiple communities.

Setting aside the information diffusion problem, there has been
some recent interest on various properties of {\em coupled} (or
{\em interacting} or {\em layered}) networks (see
\cite{ChoGohKim,KurantThiran,MarceauNoelAllard,LeichtDSouza,MendiolaSerranoBoguna}). 
For instance, \cite{ChoGohKim}, \cite{KurantThiran} and
\cite{MarceauNoelAllard} consider a layered network structure where the
networks in distinct layers are composed of {\em identical} nodes.
On the other hand, in \cite{LeichtDSouza}, 
the authors studied
the percolation problem in two interacting networks with
completely {\em disjoint} vertex sets; their model is similar to
{\em interdependent} networks introduced in \cite{Buldyrev}. 
Recently, \cite{MendiolaSerranoBoguna} studied the 
susceptible-infectious-susceptible (SIS) epidemic model
in an interdependent network.



\subsection{Summary of Main Contributions}

The current paper belongs to the second type of studies introduced above
and aims to develop a new theoretic framework towards
understanding the characteristics of information diffusion across
{\em multiple} coupled networks. Although empirical studies are 
valuable in their own right, the modeling approach adopted 
here reveals subtle relations between the network
parameters and the dynamics of information diffusion, thereby allowing 
us to develop a fundamental understanding as to how
conjoining multiple networks extends the scale of information 
diffusion. The interested reader is also referred to the article
by Epstein \cite{Epstein} which discusses many benefits of 
building and studying mathematical models;
see also \cite{KleinbergInBook}.

For illustration purposes, we give the definitions of our model
in the context of an overlaying social-physical network. 
Specifically, there is a physical information network
where information spreads amongst people
through {\em conventional} communication media (e.g.,
face-to-face communication, phone calls),
 and {\em conjoint to
this physical network}, there are online social
networks offering alternative platforms for information diffusion, such 
as Facebook, Twitter, YouTube, etc. 
In the interest of easy
exposition, we focus on the case where there exists only one
online social network along with the physical information network;
see the Appendix for an extension to the multiple 
social networks case.
We model the physical network and the social network as random
graphs with specified degree distributions \cite{newman2001random}.
We assume that each
individual in a population of size $n$ is a member of the physical network,
and becomes a member of the social network independently with a
certain 
probability. It is also assumed that information is
transmitted between two nodes (that are connected by a link in any
one of the graphs) according to the susceptible-infectious-recovered (SIR) model; 
see Section
\ref{sec:Model} for precise definitions.

Our main findings can be outlined as follows:
We show that the overlaying social-physical network exhibits a \lq\lq critical
point" above which {\em information epidemics} are possible; i.e.,
a single node can spread an item of information (a rumor, an
advertisement, a video, etc.) to a positive fraction of
individuals in the asymptotic limit. Below this critical threshold,
only {\em small} information outbreaks can occur and the fraction
of informed individuals always tends to zero.
We quantify the aforementioned critical point
in terms of the degree distributions of the networks 
and the fraction of individuals that are members of the online social network. 
Further, we compute the
probability that an information originating from an arbitrary
individual will yield an epidemic along with the resulting fraction
of individuals that are informed. Finally, in the cases where the fraction of informed individuals
tend to zero (non-epidemic state), we compute the expected {\em number} of individuals that
receive an information started from a single arbitrary node. 

These results are obtained by mapping the information diffusion
process to an equivalent bond percolation problem \cite{BroadbentHammersley}
in the conjoint social-physical network, and then 
analyzing the phase transition properties of the corresponding random graph
model. This problem is intricate since the relevant random graph model corresponds
to a {\em union} of coupled random graphs, 
and the results obtained in
\cite{newman2002spread,newman2001random} for single networks fall
short of characterizing its phase transition
properties. To overcome these difficulties, we introduce
a {\em multi-type} branching process and analyze it through 
an appropriate extension 
of the method of generating functions \cite{newman2002spread}.


To validate our analytical results,
we also  perform extensive simulation experiments on 
synthetic networks that exhibit similar characteristics
to some real-world networks. In particular, we verify our analysis on 
networks with power-law degree distributions
with exponential cut-off and on Erd\H{o}s-R\'enyi (ER) networks \cite{Bollobas};
it has been shown \cite{ClausetShaliziNewman} that many real networks,
including the Internet, exhibit power-law distributions with exponential cut-off.
We show that conjoining the networks can significantly increase
the scale of information diffusion even with only {\em one}
social network. To give a simple example, consider a physical information
network $\mathbb{W}$ and an online social network $\mathbb{F}$
that are ER graphs with respective 
mean degrees $\lambda_w$ and
$\lambda_f$, and assume that each node in $\mathbb{W}$ is a member
of $\mathbb{F}$ independently with probability $\alpha$. If
$\lambda_w=0.6$ and $\alpha=0.2$, we show that information
epidemics are possible in the overlaying social-physical network
$\mathbb{H}=\mathbb{W} \cup \mathbb{F}$ whenever $\lambda_f \geq
0.77$. In stark contrast, this happens only if $\lambda_w>1$ or
$\lambda_f>1$ when the two networks are disjoint. Furthermore, in
a single ER network $\mathbb{W}$ with $\lambda_w=1.5$, an
information item originating from an arbitrary individual gives
rise to an epidemic with probability $0.58$ (i.e., can reach at
most $58\%$ of the individuals). However, if the same network
$\mathbb{W}$ is conjoined with an ER network $\mathbb{F}$ with
$\alpha=0.5$ and $\lambda_f=1.5$, the probability of an epidemic
becomes $0.82$ (indicating that up to $82 \%$ of the population
can be influenced). These results show that the conjoint
social-physical network can spread an item of information to a
significantly larger fraction of the population as compared to the
case where the two networks are disjoint. 

The above conclusions are predicated on the social network $\mathbb{F}$
containing a {\em positive} fraction of the population. This assumption is indeed
realistic since more than 50\%
of the adult population in the US use Facebook \cite{BakshyRosennMarlowAdamic}.
However, for completeness we also analyze (see Section \ref{sec:small_social_network})
the case where the social network $\mathbb{F}$ 
contains only $\lfloor n^{\gamma} \rfloor$ nodes with $\gamma<1$.
In that case, we show analytically 
that no matter how connected $\mathbb{F}$ is, conjoining it to 
the physical network $\mathbb{W}$ does not change the threshold and the expected
size of information epidemics.

Our results provide a complete characterization
of the information diffusion process in a coupled social-physical network, by revealing the relation 
between the network parameters and the most interesting quantities including the critical threshold,
probability and expected size of information epidemics. 
To the best of our knowledge, there has been no work in the
literature that studies the information diffusion in overlay
networks whose vertices are neither identical nor disjoint.
We believe that our findings along this line shed light on the
understanding on information propagation across coupled
social-physical networks.

\subsection{Notation and Conventions}

All limiting statements, including asymptotic equivalences, are
understood with $n$ going to infinity. The random variables (rvs)
under consideration are all defined on the same probability space
$(\Omega, {\cal F}, \mathbb{P})$. Probabilistic statements are
made with respect to this probability measure $\mathbb{P}$, and we
denote the corresponding expectation operator by $\mathbb{E}$. The
mean value of a random variable $k$ is denoted by $<k>$. We use
the notation ${\buildrel st \over =}$ to indicate distributional
equality, $  \xrightarrow{a.s.}$  to indicate almost sure
convergence and $\xrightarrow{p}$ to indicate convergence in
probability. For any discrete set $S$ we write $|S|$ for its
cardinality. 
For a random graph $\mathcal{G}$ we write
$C_i(\mathcal{G})$ for the number of nodes in its $i$th largest
connected component; i.e., $C_1(\mathcal{G})$ stands for the size
of the largest component, $C_2(\mathcal{G})$ for the size of the
second largest component, etc.

The indicator function of an event $E$ is denoted by $\1{E}$. We
say that an event holds {\em with high probability} (whp) if it
holds with probability $1$ as $n \to \infty$. For sequences
$\{a_n\},\{b_n\}:
\mathbb{N}_0 \rightarrow \mathbb{R}_+$, we write 
$a_n = o(b_n)$ as a shorthand for the relation
 $\lim_{n \to \infty} \frac{a_n}{b_n}=0$, whereas $a_n = O(b_n)$
means that there exists $c>0$ such that $a_n \leq c b_n$ for all
$n$ sufficiently large. Also, we have $a_n = \Omega(b_n)$ if
$b_n=O(a_n)$, or equivalently, if there exists $c
> 0$ such that $a_n \geq c  b_n$ for all $n$ sufficiently
large. Finally, we write $a_n = \Theta(b_n)$ if we have $a_n =
O(b_n)$ and $a_n = \Omega(b_n)$ at the same time.

\subsection{Organization of the Paper}

The rest of the paper is organized as follows. In Section
\ref{sec:Model}, we introduce a model for the overlaying
social-physical network. Section \ref{sec:Results}  summarizes
the main results of the paper that deal
with the critical point and the size of information epidemics. In
section \ref{sec:Simu}, we illustrate the theoretical findings of
the paper with numerical results and verify them via extensive
simulations. In Section \ref{sec:small_social_network}, we study
information diffusion in an interesting  case where only a
sublinear fraction of individuals are members of the online social
network. The proofs of the main results are provided in
Sections \ref{sec:Proof_Arbitrary} and \ref{sec:Proof_ER}. In the
Appendix, we demonsrate an extension of the main
results to the case where there are multiple online social
networks.

\section{System Model}
\label{sec:Model}

\subsection{Overlay Network Model}
\label{sec:NetworkModel}

We consider the following model for an overlaying social-physical
network. Let $\mathbb{W}$ stand for the {\em physical} information
network of human beings on the node set $\mathcal{N}=\{1,\ldots,
n\}$. 
Next, let $\mathbb{F}$ stand for
an
online social networking web site, e.g., Facebook. We assume that
each node in $\mathcal{N}$ is a {\em member} of this auxiliary
network with probability $\alpha \in (0,1]$ independently from any
other node. In other words, we let
\begin{equation}
 \bP{i
\in \mathcal{N}_F} = \alpha, \qquad i=1,\ldots, n,
\label{eq:members_face}
\end{equation}
 with
$\mathcal{N}_F$ denoting the set of human beings that are members
of Facebook. With this assumption, it is clear that the vertex set
$\mathcal{N}_F$ of $\mathbb{F}$ satisfies
\begin{equation}
\frac{|\mathcal{N}_F|}{n} \xrightarrow{a.s.} \alpha
\label{eq:asympt}
\end{equation}
by the law of large numbers (we consider the case where
$|\mathcal{N}_F| = o(n)$ separately in Section
\ref{sec:small_social_network}).

We define the structure of the networks $\mathbb{W}$ and $\mathbb{F}$ 
through their respective degree distributions $\{p_k^w\}$ and $\{p_k^f\}$.
In particular, we specify a degree distribution that gives the
properly normalized probabilities $\{p^w_k, \: k=0,1,\ldots\}$ that
an arbitrary node in $\mathbb{W}$ has degree $k$. Then, we let
each node $i=1,\ldots, n$ in $\mathbb{W}=\mathbb{W}(n;\{p^w_k\})$
have a random degree drawn from the distribution $\{p_k^w\}$
independently from any other node. Similarly, we assume that the
degrees of all nodes in $\mathbb{F} = \mathbb{F}(n;\alpha,\{p^f_k\} )$ 
are drawn independently from
the distribution $\{p_k^f, \: k=0,1,\ldots\}$.
This corresponds to generating both networks (independently)
according to the {\em configuration} model
\cite{Bollobas,MolloyReed}. In what follows, we shall assume that the
degree distributions are well-behaved in the sense that all
moments of arbitrary order are finite.

In order to study information diffusion amongst human beings, a
key step is to characterize an overlay network $\mathbb{H}$
that is constructed by taking the union of $\mathbb{W}$ and
$\mathbb{F}$. In other words, for any distinct pair of nodes
$i,j$, we say that $i$ and $j$ are adjacent in the network
$\mathbb{H}$, denoted $i \sim_{\mathbb{H}} j$, as long as at least
one of the conditions \{$i \sim_{\mathbb{W}} j$\} or \{$i
\sim_{\mathbb{F}} j$\} holds. This is intuitive since a node $i$
can forward information to another node $j$ either by using
old-fashioned communication channels (i.e., links in $\mathbb{W}$)
or by using Facebook (i.e., links in $\mathbb{F}$). Of course, for
the latter to be possible, both $i$ and $j$ should be Facebook
users.

The overlay network $\mathbb{H}=\mathbb{W} \cup \mathbb{F}$ 
constitutes an ensemble of the {\em colored} degree-driven random 
graphs proposed in \cite{Soderberg2}. 
Let $\{1,2\}$ be the space of possible colors (or types) 
of edges in $\mathbb{H}$; specifically, we say the edges in Facebook
are of type $1$,
while the edges in the physical network are said to be of type $2$. 
The {\em colored} degree of a node $i$ is then represented
by an integer vector $\boldsymbol{d}^i=[d_f^{i},  d_w^{i}]$, 
where $d_f^{i}$ (resp. $d_w^{i}$) stands for the number of Facebook edges 
(resp. physical connections) that are incident on node $i$. 
Under the given assumptions on the 
degree distributions of $\mathbb{W}$ and $\mathbb{F}$, the colored
degrees (i.e., $\boldsymbol{d}^1, \ldots, \boldsymbol{d}^n$)
will be independent and identically distributed according to a {\em colored} degree distribution
$\{p_{\boldsymbol{d}}\}$ such that 
\begin{equation}
p_{\boldsymbol{d}} = \left(\alpha p_{d_f}^f + (1-\alpha) \1{d_f=0}\right) \cdot
p_{d_w}^w,  \quad \boldsymbol{d}=(d_f,d_w)
\label{eq:colored_dist}
\end{equation}
due to independence of $\mathbb{F}$ and $\mathbb{W}$. The term
$(1-\alpha) \1{d_f=0}$ accommodates the possibility that a 
node is not a member of the online social network, in which case the number $d_f$ of
$\mathbb{F}$-edges is automatically zero.

Given that the colored
degrees are picked such that $\sum_{i=1}^{n}d_f^{i}$ and 
$\sum_{i=1}^{n}d_w^{i}$ are even, we construct $\mathbb{H}$ as in
\cite{Soderberg2,newman2002spread}: 
Each node $i=1,\ldots,n$ is first
given the appropriate number $d_f^{i}$ and $d_w^i$ of stubs of type $1$
and type $2$, respectively. Then, pairs of these stubs that are of the
same type are picked randomly and connected together to form
complete edges; clearly, two stubs can be
connected together {\em only} if they are of the same type.
Pairing of stubs continues until none is left.

\subsection{Information Propagation Model}

Now, consider the diffusion of a piece of information in the overlay network $\mathbb{H}$
which starts from a single node.
We assume that information spreads from a node to its neighbors according
to the SIR epidemic model. In this context, an
individual is either {\em susceptible} (S) meaning that she has not
yet received a particular item of information, or {\em infectious} (I) meaning
that she is aware of the information and is capable of spreading it to
her contacts, or {\em recovered} (R) meaning that she is no
longer spreading the information. This analogy between the spread of diseases
and spread of information in a network has long been recognized \cite{Blogspace} and
SIR epidemic model is commonly used in similar studies; e.g., see \cite{Worms}
(diffusion of worms in online social networks), \cite{Blogspace} (diffusion of information 
through Blogs), and \cite{P2P} (diffusion of files in 
peer-to-peer file sharing networks), among others.

The dynamics of information diffusion can now be described as in
\cite{newman2002spread}:
We assume that an infectious individual $i$ transmits the
information to a susceptible contact $j$ with probability $T_{ij}$
where
\begin{equation}\nonumber
\label{eq:transmissibility} T_{ij} = 1-e^{-r_{ij}\tau_i}.
\end{equation}
Here, $r_{ij}$ denotes the average rate of being in contact over
the link from $i$ to $j$, and $\tau_i$ is the time $i$ keeps
spreading the information; i.e., the time it takes for $i$ to
become recovered.

It is expected that the information propagates over the physical
and social networks at different speeds, which manifests from
different probabilities $T_{ij}$ across links in this case.
Specifically, let $T^w_{ij}$ stand for the probability of
information transmission over a link (between and $i$ and $j$) in
$\mathbb{W}$ and let $T^f_{ij}$ denote the probability of
information transmission over a link in $\mathbb{F}$. For
simplicity, we assume that $T^w_{ij}$ and $T^f_{ij}$ are
independent for all distinct pairs $i,j=1, \ldots, n$.
Furthermore, we assume that the random variables $r^{w}_{ij}$ and
$\tau^{w}_{i}$ are independent and identically distributed
(i.i.d.) with probability densities $P_w(r)$ and $P_w(\tau)$,
respectively. In that case, it was shown in \cite{newman2002spread,Grassberger} 
that
information propagates over $\mathbb{W}$ as if all
transmission probabilities were equal to $T_w$, where $T_w$ is the
mean value of $T^{w}_{ij}$; i.e.,
\[
T_w:= <T^{w}_{ij}> = 1 - \int_{0}^{\infty}\int_{0}^{\infty} e^{-r
\tau}P_w(r)P_w(\tau) dr d\tau.
\]
We refer to $T_w$ as the {\em transmissibility} of the information
over the physical network $\mathbb{W}$ and note that $0 \leq T_w \leq 1$. 
In the same manner, we assume that $r^{f}_{ij}$ and
$\tau^{f}_{i}$ are i.i.d. with respective densities $P_f(r)$ and
$P_f(\tau)$ leading to a transmissibility $T_f$ of information
over the online social network $\mathbb{F}$.

Under these assumptions, information diffusion becomes 
equivalent to the {\em bond percolation} on the conjoint network
$\mathbb{H}=\mathbb{W} \cup \mathbb{F}$ \cite{newman2002spread,Grassberger}. 
More specifically, assume that each edge 
in $\mathbb{W}$ (resp. $\mathbb{F}$) is {\em
occupied} -- meaning that it can be used in spreading the
information -- with probability $T_w$ (resp. $T_f$) independently from all other
edges. Then, the size of an
information outbreak started from an arbitrary node 
is equal to the number of
individuals that can be reached from that initial node by using
only the {\em occupied} links of $\mathbb{H}$. Hence, the
threshold and the size of information epidemics can be computed
by studying the {\em phase transition} properties of the random graph 
$\mathbb{H}(n;\alpha,\{p^w_k\},T_w,\{p^f_k\},T_f) =
\mathbb{W}(n;\{p^w_k\},T_w) \cup \mathbb{F}(n;\alpha,
\{p^f_k\},T_f)$ which is obtained by taking a union of the
occupied edges of $\mathbb{W}$ and $\mathbb{F}$. More precisely,
information epidemics can take place if and only if 
$\mathbb{H}(n;\alpha,\{p^w_k\},T_w,\{p^f_k\},T_f)$ has a {\em giant} connected component
that contains a positive fraction of nodes in the large $n$ limit. Also, an arbitrary node
can trigger an information epidemic only if it belongs to the giant component, in which case 
an information started from that node will reach to all nodes in the giant component.
Hence, the fractional size of the giant component in
$\mathbb{H}(n;\alpha,\{p^w_k\},T_w,\{p^f_k\},T_f)$ gives both the probability 
that an arbitrary node triggers an information epidemic as well as the corresponding 
fractional size of the information epidemic. 

\section{Main Results}
\label{sec:Results}


\subsection{Information Diffusion in Coupled Graphs with Arbitrary Degree Distributions}
\label{sec:Results_Arbitrary}

We now present the main result of our paper characterizes the
threshold and the size of the information epidemic in $\mathbb{H}$
by revealing its phase transition properties. First, for
notational convenience, let $k_f$ and $k_w$ be random variables
independently drawn from the distributions $\{p_k^f\}$ and
$\{p_k^w\}$, respectively, and let $<k_f> := \lambda_f$ and $<k_w>
:= \lambda_w$. Further, assume that $\beta_f$ and $\beta_w$
are given by
\begin{equation}
\label{eq:betas_new_def}
\beta_f:= \frac{<k_f^2> -\lambda_f }{\lambda_f} \quad \textrm{and}
\quad \beta_w:= \frac{<k_w^2> -\lambda_w }{\lambda_w},
\end{equation}
and define the threshold function $\sigma_{fw}^{\star}$ by
\begin{eqnarray}\label{eq:sigma_star}
\lefteqn{\sigma_{fw}^{\star}} &&
\\ \nonumber
&=& \frac{T_f \beta_f + T_w\beta_w + \sqrt{(T_f\beta_f -
T_w\beta_w)^2 + 4\alpha T_f T_w \lambda_f \lambda_w}}{2}
\end{eqnarray}
Finally, let $h_1,h_2$ in $(0,1]$ be given by the pointwise
smallest solution of the recursive equations
\begin{eqnarray}
h_1 = \frac{T_f}{\lambda_f}
\bE{k_f h_1^{k_f-1}}
\label{eq:h_1} 
 \bE{h_2^{k_w}} + 1-T_f  \hspace{1.7cm}\\
h_2 \label{eq:h_2} = \frac{T_w}{\lambda_w} \bE{\alpha
h_1 ^{k_f} + 1-\alpha}  
\bE{k_w h_2^{k_w-1}} + 1-T_w. 
\end{eqnarray}

\begin{thm}
{\sl Under the assumptions just stated, we have
\begin{itemize}
\item[$(i)$] If $\sigma_{fw}^{\star} \leq 1$ then with high
probability the size of the largest component satisfies
$C_1\left(\mathbb{H} (n;\alpha, \{p_k^w\},T_w,
\{p_k^f\},T_f)\right)=o(n)$. On the other hand, if
$\sigma_{fw}^{\star}
> 1$, then $C_1\left(\mathbb{H} (n;\alpha, \{p_k^w\},T_w,
\{p_k^f\},T_f)\right)=\Theta(n)$ whp. \item[$(ii)$] Also,
\begin{eqnarray}
\lefteqn{\frac{1}{n} C_1\left(\mathbb{H} (n;\alpha,
\{{p}_k^w\}, T_w,
\{{p}_k^f\},T_f)\right) } &&\nonumber \\
 &\xrightarrow{p}& 1- \bE{\alpha h_1^{k_f} +
1-\alpha}   \bE{h_2^{k_w}} .
\label{eq:part2_main_arbitrary}
\end{eqnarray}
\end{itemize}
} \label{thm:main_Arbitrary}
\end{thm}

A proof of Theorem \ref{thm:main_Arbitrary} is given in Section
\ref{sec:Proof_Arbitrary}.

Theorem \ref{thm:main_Arbitrary}  
quantifies the fraction of individuals in
the overlaying social-physical network that are likely to receive
an item of information which starts spreading from a single
individual. 
Specifically, Theorem \ref{thm:main_Arbitrary} shows that the
critical point of the information epidemic is marked by
$\sigma_{fw}^{\star} = 1$, with the critical threshold
$\sigma_{fw}^{\star}$ given by (\ref{eq:sigma_star}). In other
words, for any parameter set that yields $\sigma_{fw}^{\star}
> 1$ (supercritical regime), an item of information has a positive probability
of giving rise to an information epidemic; i.e., reaching a
linear fraction of the individuals. In that case, the probability of 
a node triggering an information epidemic, and the
corresponding asymptotic fraction of individuals who receive the information can be
found by first solving the recursive equations
(\ref{eq:h_1})-(\ref{eq:h_2}) for the smallest $h_1,h_2$ in
$(0,1]$ and then computing the expression given in
(\ref{eq:part2_main_arbitrary}). On the other hand, whenever it
holds that $\sigma_{fw}^{\star} \leq 1$ (subcritical regime), we
conclude from Theorem \ref{thm:main_Arbitrary} that the
number of individuals who receive the information will be $o(n)$
with high probability, meaning that all information outbreaks are
non-epidemic.

It is of interest to state whether or not Theorem
\ref{thm:main_Arbitrary} can be deduced from the phase transition results
for random graphs with arbitrary degree distributions (e.g., see
 \cite{MolloyReed,newman2002spread,newman2001random}). It is
 well known \cite{MolloyReed} that for these graphs the critical point of the phase
 transition is given by
 \[
\frac{\bE{d_i(d_i-1)}}{\bE{d_i}} =1
 \]
where $d_i$ is the degree of an arbitrary node. We next show that
this condition is not equivalent (and, indeed is not even a good
approximation) to  $\sigma_{fw}^{\star}=1$.

To this end, we consider
a basic scenario where $\mathbb{F}$ and $\mathbb{W}$ are both
Erd\H{o}s-R\'{e}nyi graphs \cite{Bollobas} so that their degree distributions are 
(asymptotically) Poisson, i.e., we have $p_k^w= e^{-\lambda_w}
\frac{\lambda_w^k}{k!}$ and $p_k^f= e^{-\lambda_f}
\frac{\lambda_f^k}{k!}$. Given that each link in $\mathbb{F}$ (resp. in $\mathbb{W}$)
is occupied with probability $T_f$ (resp. $T_w$), 
the occupied degree of an arbitrary node $i$ in $\mathbb{H}$ follows
a Poisson distribution with mean $T_w\lambda_w$ if $i \not \in
\mathcal{N}_F$ (which happens with probability $1-\alpha$), and it
follows a Poisson distribution with mean
$T_f\lambda_f+T_w\lambda_w -\frac{T_f\lambda_f T_w\lambda_w}{n}$
if $i \in \mathcal{N}_F$ (which happens with probability $\alpha$). When
$n$ becomes large this leads to
\begin{equation}
\frac{\bE{d_i(d_i-1)}}{\bE{d_i}} = \frac{\alpha
(T_f\lambda_f+T_w\lambda_w)^2 + (1-\alpha)
(T_w\lambda_w)^2}{\alpha T_f\lambda_f + T_w\lambda_w}.
\label{eq:threshold_candidate}
\end{equation}
It can be seen that the above expression is not equal to
the corresponding quantity $\sigma_{fw}^{\star}$ -- As discussed in
the next subsection, for the given degree
distributions we have $\sigma_{fw}^{\star}=\lambda_{fw}^{\star}$, where
$\lambda_{fw}^{\star}$ is given by (\ref{eq:lambda_fw_star_first}).
For instance, with $\alpha=0.2$, $T_w\lambda_w=0.6$ and
$T_f\lambda_f=0.8$, we have $\sigma_{fw}^{\star} = \lambda_{fw}^{\star} = 1.03$ 
while
(\ref{eq:threshold_candidate}) yields $0.89$ signaling a
significant difference between the exact threshold
$\lambda_{fw}^{\star}$ and the approximation given by
(\ref{eq:threshold_candidate}). We conclude that the results
established in Theorem \ref{thm:main_Arbitrary} (for coupled random graphs)
go beyond  the classical results for {\em single} random
graphs with arbitrary degree distributions.

Aside from the critical threshold and the fractional size of information epidemics, 
we are also interested in computing the {\em average} size of information outbreaks
in the subcritical regime for a fuller understanding of information propagation process. 
In other words, in the case where the fraction of informed individuals tends to 
zero, we wish to compute the expected {\em number} of informed nodes.
For a given network with nodes $1, \ldots, n$, 
the average outbreak size $<s>$  is given by $\sum_{i=1}^{n}\frac{1}{n} s(i)$, where
$s(i)$ is the number of nodes that receive an information started from node $i$; i.e., $s(i)$
is the size of the largest connected component containing node $i$.

Now, let $<s>:=<s(n;\alpha, \{p_k^w\},T_w,
\{p_k^f\},T_f))>$ denote the average outbreak size in $\mathbb{H}(n;\alpha, \{p_k^w\},T_w,
\{p_k^f\},T_f)$. It is easy to check that 
\begin{equation}
<s> = \sum_{j=1}^{N_{c}} \frac{1}{n} \left(C_j(\mathbb{H} (n;\alpha, \{p_k^w\},T_w,
\{p_k^f\},T_f))\right)^2
\label{eq:average_outbreak}
\end{equation}
where, as before, $C_j$ gives the size of the $j$th largest component of the network,
and $N_c$ denotes the total number of components.
To see (\ref{eq:average_outbreak}), observe that an arbitrarily selected node will belong
to a component of size $C_j$ with probability $C_j/n$, in which case an information started from
that particular node will create an outbreak of size $C_j$. Summing over all 
components of the network, we get (\ref{eq:average_outbreak}). 
In the supercritical regime, we have $C_1(\mathbb{H}) = \Omega(n)$ so that 
$<s> \to \infty$. The next result, established in Section \ref{sec:Proof_Arbitrary}, 
allows computing this quantity in
the subcritical regime.
\begin{thm}
{\sl Let $\sigma_{fw}^{\star} \leq 1$. With the above assumptions,
let $s_1$, $s_2$ denote the simultaneous stable solution
of the equations
\begin{eqnarray}\label{eq:s_1}
s_1&=& T_f +\beta_f T_f s_1 + \lambda_w T_f s_2 \\ 
s_2 &=& T_w+ \alpha \lambda_f T_w s_1 + \beta_w T_w s_2
\label{eq:s_2}
\end{eqnarray}
Then, the average outbreak size satisfies 
\begin{equation}
<s>~\xrightarrow{p}~1+ \alpha \lambda_f s_1 + \lambda_w s_2.
\end{equation}
} \label{thm:main_average_size}
\end{thm}

\subsection{Special Case: Information Diffusion in coupled ER graphs}
\label{sec:Results_ER}

A special case of interest is when both $\mathbb{W}$ and
$\mathbb{F}$ are Erd\H{o}s-R\'{e}nyi graphs \cite{Bollobas}. More
specifically, let $\mathbb{W}=\mathbb{W}(n;\lambda_w/n)$ be an ER
network on the vertices $\{1,\ldots,n\}$ such that there exists an
edge between any pair of distinct nodes $i,j=1,\ldots,n$ with probability
$\lambda_w/n$; this ensures that mean degree of each node is
asymptotically equal to $\lambda_w$. Next, obtain a set of
vertices $\mathcal{N}_F$ by picking each node $1,\ldots, n$
independently with probability $\alpha \in (0,1]$. Now, let
$\mathbb{F}=\mathbb{F}(n; \alpha, \lambda_f/(\alpha n))$ be an ER
graph on the vertex set $\mathcal{N}_F$ with edge probability
given by $\frac{\lambda_f}{\alpha n}$. The mean degree of a node
in $\mathbb{F}$ is given (asymptotically) by $\lambda_f$ as seen
via (\ref{eq:asympt}). 

Given that the degree distributions are asymptotically Poisson in ER graphs,
this special case is covered by
our model presented in Section \ref{sec:NetworkModel} by setting 
$p_k^w= e^{-\lambda_w}
\frac{\lambda_w^k}{k!}$ and $p_k^f= e^{-\lambda_f}
\frac{\lambda_f^k}{k!}$. Thus, Theorem \ref{thm:main_Arbitrary} is still valid
and can be used to obtain the condition and expected size of information epidemics.  
However, recent developments on inhomogeneous random graphs
 \cite{BollobasJansonRiordan} enable us to obtain more detailed results than those 
 given by Theorem \ref{thm:main_Arbitrary} for this special case.

Consider now an overlay network model $\mathbb{H}$ 
constructed on the vertices $1,\ldots,n$ by conjoining the {\em
occupied} edges of $\mathbb{W}$ and $\mathbb{F}$, i.e., we have
$\mathbb{H}(n;\alpha, T_w \lambda_w,  T_f \lambda_f ) =
\mathbb{W}(n;T_w\lambda_w/n) \cup
\mathbb{F}(n;\alpha,T_f\lambda_f/(\alpha n))$. 
Let $\lambda_{fw}^{\star}$ be defined by
\begin{eqnarray}\label{eq:lambda_fw_star_first}
\lambda_{fw}^{\star} &:=&
\frac{1}{2} \left(T_f\lambda_f+T_w\lambda_w\right)  \\
& & ~ + \frac{1}{2} \sqrt{\left(T_f\lambda_f+T_w\lambda_w
\right)^2-4(1-\alpha)T_f\lambda_f T_w\lambda_w}. \nonumber
\end{eqnarray}
Also, let $\rho_1, \rho_2$ be the pointwise largest solution of the recursive equations
\begin{equation}\begin{array}{l}
  \rho_1 = 1-\exp\left\{- \rho_1(\alpha \lambda_wT_w +\lambda_f T_f)-\rho_2 ( 1 - \alpha ) \lambda_w T_w \right\} \\
 \rho_2 = 1-\exp\left\{ -\rho_1 \alpha\lambda_w T_w -  \rho_2 ( 1 - \alpha ) \lambda_w T_w \right\} \\
\end{array}
\label{eq:ro_1_ER_first}
\end{equation}
with $\rho_1, \rho_2$ in $[0,1]$.

\begin{thm}
{\sl With the above assumptions, we have
\begin{itemize}
\item[$(i)$] If $\lambda_{fw}^{\star} \leq 1$, then with high
probability, the size of the largest component satisfies
$C_1(\mathbb{H}(n;\alpha, T_w \lambda_w, T_f \lambda_f))=O(\log
n)$; in contrast, if $\lambda_{fw}^{\star}  > 1$ we have
$C_1(\mathbb{H}(n;\alpha,T_w\lambda_w,T_f\lambda_f))=\Theta(n)$
whp, while the size of the second largest component satisfies
$C_2(\mathbb{H}(n;\alpha, T_w \lambda_w, T_f \lambda_f))=O(\log
n)$. \item[$(ii)$] Moreover,
\begin{equation}\nonumber
\frac{1}{n} C_1(\mathbb{H}(n;\alpha, T_w \lambda_w, T_f
\lambda_f)) \xrightarrow{p} \alpha \rho_1 + (1-\alpha) \rho_2.
\end{equation}
\end{itemize}
} \label{thm:main_ER}
\end{thm}

A proof of Theorem \ref{thm:main_ER} is given in Section
\ref{sec:Proof_ER}

Theorem \ref{thm:main_ER}  is a counter-part of 
Theorem \ref{thm:main_Arbitrary}. This time,
the \lq \lq critical point" of the
information epidemic is marked by $\lambda_{fw}^{\star} = 1$, with
the critical threshold $\lambda_{fw}^{\star}$ given by
(\ref{eq:lambda_fw_star_first}).  With $p_k^w= e^{-\lambda_w}
\lambda_w^k/{k!}$ and $p_k^f= e^{-\lambda_f}
\lambda_f^k/{k!}$, we have that $\beta_f = \lambda_f$,
$\beta_w = \lambda_w$, and  it is easy to check that
$\sigma_{fw}^{\star} = \lambda_{fw}^{\star}$ so that part $(i)$ of
Theorem \ref{thm:main_ER} is compatible with part $(i)$ of
Theorem \ref{thm:main_Arbitrary}. Also, we find (numerically) that the
second parts of Theorems \ref{thm:main_ER} and 
\ref{thm:main_Arbitrary} yield the same asymptotic giant component
size. Nevertheless, it is worth noting that Theorem
\ref{thm:main_ER} is not a corollary of Theorem
\ref{thm:main_Arbitrary}. This is because, through a different
technique used in the proofs, Theorem \ref{thm:main_ER} provides the
sharper bounds $C_1(\mathbb{H} (n;\alpha, T_w \lambda_w, T_f
\lambda_f)=O(\log n)$ (subcritical case) and $C_2(\mathbb{H}
(n;\alpha, T_w \lambda_w, T_f \lambda_f)=O(\log n)$ (supercritical
case) that go beyond Theorem \ref{thm:main_Arbitrary}.

\begin{figure}[!t]
\vspace{-2mm}
\hspace{-4mm}
\includegraphics[totalheight=0.33\textheight,
width=.55\textwidth]{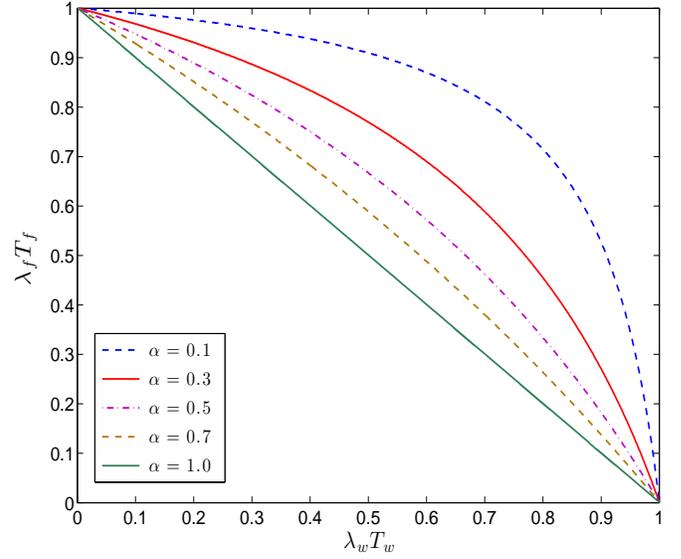}\caption{\sl The minimum
$\lambda_f T_f$ required for existence of a giant component in
$\mathbb{H}(n;\alpha,T_w\lambda_w,T_f\lambda_f)$ versus $\lambda_w
T_w$ for various $\alpha$ values. In other words, each curve
corresponds to the boundary of the phase transition for the
corresponding $\alpha$ value. Above the boundary there exists a
giant component, but below it all components have $O(\log n)$
nodes. } \label{fig:boundary}
\end{figure}

We observe that the threshold function $\lambda_{fw}^{\star}$ is
symmetric in $T_f \lambda_f $ and $T_w \lambda_w$, meaning that
both networks have identical roles in carrying the conjoined network to the
supercritical regime where information can reach a linear
fraction of the nodes. To get a more concrete sense,
 we depict in Figure \ref{fig:boundary} the minimum
$\lambda_f T_f$ required to have a giant component in
$\mathbb{H}(n;\alpha, T_w\lambda_w, T_f \lambda_f)$ versus
$\lambda_w T_w$ for various $\alpha$ values. Each curve in the
figure corresponds to a phase transition boundary above which
information epidemics are possible. If $T_f=T_w=1$, the same plot
shows the boundary of the giant component existence with respect
to the mean degrees $\lambda_f$ and $\lambda_w$. This clearly
shows how two networks that are in the subcritical regime can
yield an information epidemic when they are conjoined.
For instance, we see that for $\alpha=0.1$, it suffices to have
$\lambda_f = \lambda_w = 0.76 $ for the existence of an
information epidemic. Yet, if the two networks were disjoint, it
would be necessary \cite{Bollobas} to have $\lambda_f>1$ and
$\lambda_w>1$. 

We elaborate further on Theorem \ref{thm:main_ER}. First, we note
from the classical results \cite{Bollobas} that ER graphs have a
giant component whenever average node degree exceeds one. This is
compatible with part $(i)$ of Theorem \ref{thm:main_ER}, since the
condition for giant component existence reduces to $T_f \lambda_f
> 1$ if $T_w\lambda_w=0$ and $T_w \lambda_w>1$ when $T_f\lambda_f = 0$.
Finally, in the case where $\alpha=1$ (i.e., when
everyone in the population is a member of Facebook), the graph
$\mathbb{H}$ reduces to an ER graph with edge probability
$\frac{T_f \lambda_f+ T_w\lambda_w - \frac{T_f T_w
\lambda_f\lambda_w}{n}}{n}$ leading to a mean node degree of $T_f
\lambda_f + T_w \lambda_w$ in the asymptotic regime. As expected,
for the case $\alpha=1$, Theorem \ref{thm:main_ER} reduces to
classical results for ER graphs as we see that
$\lambda_{fw}^{\star} = T_f \lambda_f + T_w \lambda_w$ and
$\frac{1}{n} C_1(\mathbb{H}) \xrightarrow{p}   \rho_1$ where
$\rho_1$ is the largest solution of
$
\rho_1 = 1- e^{-\rho_1(T_f\lambda_f+T_w\lambda_w)}.
$

\section{Numerical Results}
\label{sec:Simu}


\subsection{ER Networks}
\label{sec:Simu_ER}

We first study the case where both the physical information
network $\mathbb{W}$ and the online social network $\mathbb{F}$
are Erd\H{o}s-R\'enyi graphs. As in Section \ref{sec:Results_ER},
let $\mathbb{H}(n;\alpha, T_w \lambda_w,  T_f \lambda_f ) =
\mathbb{W}(n;T_w\lambda_w/n) \cup
\mathbb{F}(n;\alpha,T_f\lambda_f/(\alpha n))$ be the conjoint
social-physical network, where $\mathbb{W}$ is defined on the
vertices $\{1, \ldots, n\}$, whereas the vertex set of
$\mathbb{F}$ is obtained by picking each node $1,\ldots, n$
independently with probability $\alpha$. The information
transmissibilities are equal to $T_w$ and $T_f$
in $\mathbb{W}$ and $\mathbb{F}$, respectively, so that 
the mean degrees are given (asymptotically) by
$T_w \lambda_w $ and $ T_f \lambda_f$, respectively.

We plot in Figure \ref{fig:giant_size_ER} the fractional size of
the giant component in $\mathbb{H}(n;\alpha, T_w \lambda_w,  T_f
\lambda_f )$ versus $T_f \lambda_f = T_w \lambda_w$ for various
$\alpha$ values. In other words, the plots illustrate the largest
fraction of individuals that a particular item of information can reach.
In this figure, the curves stand for the analytical results
obtained by Theorem \ref{thm:main_ER} whereas marked points stand
for the experimental results obtained with $n=2 \times 10^5$ nodes by
averaging $200$ experiments for each data point. Clearly, there
is an excellent match between the theoretical and experimental results.
It is also seen that the critical
threshold for the existence of a giant component (i.e., an
information epidemic) is given by $T_f \lambda_f = T_w \lambda_w=
0.760$ when $\alpha=0.1$, $T_f \lambda_f = T_w \lambda_w= 0.586$
when $\alpha=0.5$, and $T_f \lambda_f = T_w \lambda_w= 0.514$ when
$\alpha=0.9$. It is easy to check that these values are in perfect
agreement with the theoretically obtained critical threshold
$\lambda_{fw}^{\star}$ given by (\ref{eq:lambda_fw_star_first}). 

In the inset of Figure \ref{fig:giant_size_ER}, we demonstrate 
the average outbreak size $<s>$ versus $T_f \lambda_f = T_w \lambda_w$
under the same setting. Namely, the curves
stand for the analytical results obtained from Theorem \ref{thm:main_average_size},
while the marked points are obtained by averaging 
the quantity given in (\ref{eq:average_outbreak})
over 200 independent experiments. We see that 
experimental results are in excellent agreement with our analytical results. 
Also, as expected, average outbreak size $<s>$ is seen to grow unboundedly as 
$T_f \lambda_f = T_w \lambda_w$ approaches to the corresponding epidemic threshold.

\begin{figure}[!t]
\hspace{-4mm}
\includegraphics[totalheight=0.33\textheight,
width=.55\textwidth]{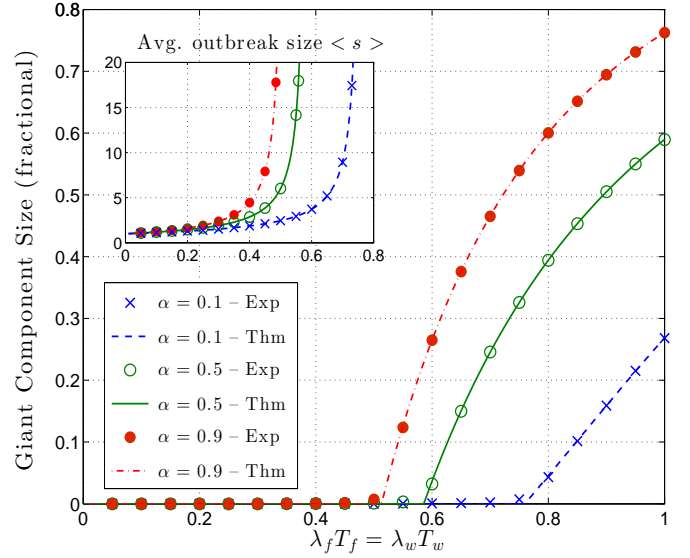}\caption{\sl The fractional
size of the giant component in $\mathbb{H}(n;\alpha,
T_w\lambda_w,T_f \lambda_f)$ versus $T_f \lambda_f = T_w
\lambda_w$. The curves correspond to analytical results obtained
from Theorem \ref{thm:main_ER}, whereas marked points stand for
the experimental results obtained with $n=2\times 10^5$ by averaging
$200$ experiments for each point. (Inset) Average out-break size $<s>$ 
versus $T_f \lambda_f = T_w \lambda_w$ under the same setting.
} \label{fig:giant_size_ER}
\end{figure}

\subsection{Networks with Power Degree Distributions}
\label{sec:Simu_Power}

 In order to gain more insight about the consequences of 
 Theorem \ref{thm:main_Arbitrary} for real-world networks,
we now consider a specific example of information diffusion when the
physical information network $\mathbb{W}$ and the online social
network $\mathbb{F}$ have power-law degree distributions with exponential cutoff.
%
Specifically, we let
\begin{equation}
p_k^w = \left\{
\begin{array}{ll}
  0  & \textrm{if $k=0$}\\
 \left(\mbox{Li}_{\gamma_w}(e^{-1/\Gamma_w})\right)^{-1} k^{-\gamma_w} e^{-k/\Gamma_w} & \textrm{if $k=1,2,\ldots$} \\
\end{array}
\right. \label{eq:p_k_w}
\end{equation}
and
\begin{equation}\label{eq:p_k_f}
p_k^f = \left\{
\begin{array}{ll}
  0  & \textrm{if $k=0$}\\
 \left(\mbox{Li}_{\gamma_f}(e^{-1/\Gamma_f})\right)^{-1} k^{-\gamma_f} e^{-k/\Gamma_f} & \textrm{if $k=1,2,\ldots$},\\
\end{array}
\right.
\end{equation}
where $\gamma_w$, $\gamma_f$, $\Gamma_w$ and $\Gamma_f$ are
positive constants and the normalizing constant $\mbox{Li}_{m}(z)$
is the $m$th polylogarithm of $z$; i.e.,
$
\mbox{Li}_{m}(z) = \sum_{k=1}^{\infty} \frac{z^k}{k^m}.
$

Power law
distributions with exponential cutoff are chosen here because they
are  applied to a variety of real-world networks \cite{newman2002spread,LeichtDSouza}. 
In fact, a detailed
empirical study on the
degree distributions of real-world networks
\cite{ClausetShaliziNewman} 
revealed that the {\em Internet} (at the level of autonomous systems), 
the {\em phone call} network, the {\em e-mail} network, and the {\em web link} network
all exhibit power law degree distributions with exponential cutoff.


To apply Theorem \ref{thm:main_Arbitrary}, we first compute the
epidemic threshold given by (\ref{eq:sigma_star}). Under
(\ref{eq:p_k_w})-(\ref{eq:p_k_f}) we find that
\begin{eqnarray}
\lambda_f &=&
\frac{\mbox{Li}_{\gamma_f-1}(e^{-1/\Gamma_f})}{\mbox{Li}_{\gamma_f}(e^{-1/\Gamma_f})},
\nonumber \\
\beta_f &=& \frac{\mbox{Li}_{\gamma_f-2}(e^{-1/\Gamma_f}) -
\mbox{Li}_{\gamma_f-1}(e^{-1/\Gamma_f})}{\mbox{Li}_{\gamma_f-1}(e^{-1/\Gamma_f})}
\nonumber
\end{eqnarray}
Similar expressions can be derived for $\lambda_w$ and $\beta_w$.
It is now a simple matter to compute the critical threshold
$\sigma_{fw}^{\star}$ from (\ref{eq:sigma_star}) using the above
relations. Then,  we can use Theorem \ref{thm:main_Arbitrary}(i)
to check whether or not an item of information can reach a linear
fraction of individuals in the conjoint social-physical network
$\mathbb{H}(n;\alpha,\{p_k^w\},T_w, \{p_k^f\}, T_f) =
\mathbb{W}(n;\{p_k^w\},T_w) \cup
\mathbb{F}(n;\alpha,\{p_k^f\},T_f)$.

To that end, we depict in Figure \ref{fig:boundary2} the minimum
$T_f$ value required to have a giant component in
$\mathbb{H}(n;\alpha, \{p_k^w\}, T_w, \{p_k^f\}, T_f )$
versus $T_w$, for various $\alpha$ values. In other words, each
curve corresponds to a phase transition boundary above which
information epidemics are possible, in the sense that an
information has a positive probability of reaching out to a linear
fraction of individuals in the overlaying social-physical network.
In all plots, we set $\gamma_f=\gamma_w = 2.5$ and
$\Gamma_f=\Gamma_w=10$. The $T_f$ and $T_w$ values are multiplied
by the corresponding $\beta_f$ and $\beta_w$ values to make a fair
comparison with the disjoint network case where it is required
\cite{newman2002spread} to have $\beta_w T_w > 1$ (or $\beta_f T_f
> 1$) for the existence of an epidemic; under the current
setting we have $\beta_f=\beta_w = 1.545$. Figure
\ref{fig:boundary2} illustrates how conjoining two networks can 
speed up the information
diffusion. It can be seen that even for small $\alpha$
values, two networks, albeit having no giant component
individually, can yield an information epidemic when they are
conjoined. As an example, we see that for
$\alpha=0.1$, it suffices to have that $\beta_f T_f =\beta_w T_w=
0.774 $ for the existence of an information epidemic in the
conjoint network $\mathbb{H}$, whereas if the networks
$\mathbb{W}$ and $\mathbb{F}$ are disjoint, an information
epidemic can occur only if $\beta_w T_w
> 1$ or $\beta_f T_f> 1$.

\begin{figure}[!t]
\hspace{-5mm}
\includegraphics[totalheight=0.33\textheight,
width=.55\textwidth]{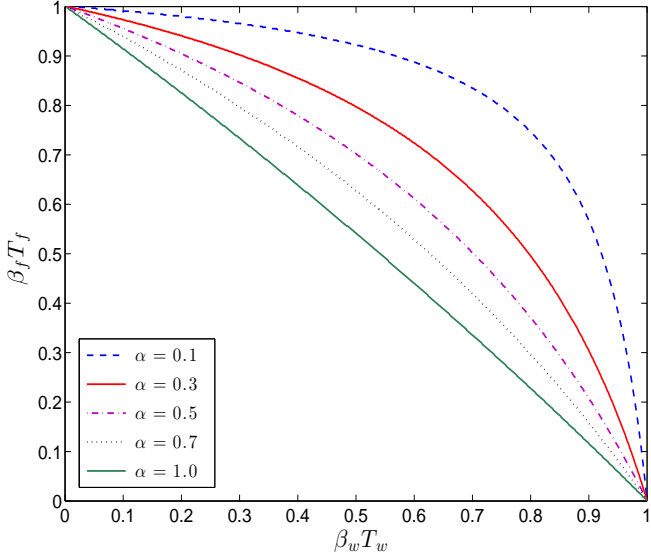}\caption{\sl The minimum $T_f$
required for the existence of a giant component in
$\mathbb{H}(n;\alpha,\{p_k^w\},T_w,\{p_k^f\},T_f)$ versus $T_w$.
 The distributions $\{p_k^w\}$ and $\{p_k^f\}$ are given by (\ref{eq:p_k_w}) and (\ref{eq:p_k_f}),
 with $\gamma_f=\gamma_w = 2.5$ and
$\Gamma_f=\Gamma_w=10$. The $T_f$ and $T_w$ values are multiplied
by the corresponding $\beta_f$ and $\beta_w$ values to provide a fair
comparison with the disjoint network case; under the current
setting we have $\beta_f=\beta_w = 1.545$. } \label{fig:boundary2}
\end{figure}

\begin{figure}[!t]
\hspace{-4mm}
\includegraphics[totalheight=0.33\textheight,
width=.55\textwidth]{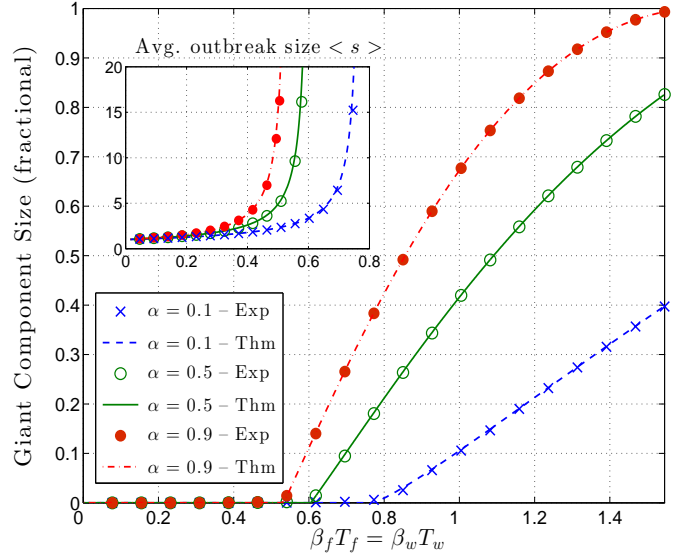}\caption{\sl The
fractional size of the giant component in $\mathbb{H}(n;\alpha,
\{p_k^w\},T_w, \{p_k^f\},T_f)$ versus $T_f \beta_f = T_w \beta_w$.
 The distributions $\{p_k^w\}$ and
$\{p_k^f\}$ are given by (\ref{eq:p_k_w}) and (\ref{eq:p_k_f}),
 with $\gamma_f=\gamma_w = 2.5$ and
$\Gamma_f=\Gamma_w=10$. The $T_f$ and $T_w$ values are multiplied
by the corresponding $\beta_f$ and $\beta_w$ values for fair
comparison with the disjoint network case; under the current
setting we have $\beta_f=\beta_w = 1.545$. The curves were obtained analytically via Theorem
\ref{thm:main_Arbitrary}, whereas the marked points stand for the
experimental results obtained with $n=2 \times 10^5$ nodes by averaging $200$
experiments for each parameter set. We see that there is an excellent
agreement between theory and experiments. (Inset) Average out-break size $<s>$ 
versus $\beta_f T_f = \beta_w T_w$ under the same setting.}
\label{fig:giant_size_Power}
\end{figure}

Next, we turn to computation of the giant component size. We
note that
\[
\begin{array}{c}
\bE{h_1^{k_f}}=
\frac{\mbox{Li}_{\gamma_f}(h_1 e^{-1/\Gamma_f})}{\mbox{Li}_{\gamma_f}(e^{-1/\Gamma_f})} \hspace{-3mm}
\\
\bE{k_f h_1^{k_f-1}} 
=\frac{\mbox{Li}_{\gamma_f-1}(h_1e^{-1/\Gamma_f})}{\mbox{Li}_{\gamma_f}(e^{-1/\Gamma_f})h_1}.
\end{array}
\]
and similar expressions can be derived for
$\mathbb{E}[h_2^{k_w}]$ and $\mathbb{E}[k_w h_2^{k_w-1}] $.
Now, for any given set of parameters,
$\gamma_f,\gamma_w,T_f,T_w,\Gamma_f,\Gamma_w,\alpha$, we can
numerically obtain the giant component size of
$\mathbb{H}(n;\alpha,\{p_k^w\},T_w, \{p_k^f\}, T_f)$ by
invoking the above relations into part $(ii)$ of Theorem
\ref{thm:main_Arbitrary}.

To this end, Figure \ref{fig:giant_size_Power} depicts the
fractional size of the giant component in $\mathbb{H}(n;\alpha,
\{p_k^w\},T_w, \{p_k^f\},T_f)$ versus $T_f \beta_f = T_w \beta_w$,
for various $\alpha$ values; as before, we set
$\gamma_f=\gamma_w=2.5$ and $\Gamma_f=\Gamma_w=10$ yielding
$\beta_f=\beta_w = 1.545$. In other words, the plots stand for the
largest fraction of individuals in the social-physical network who
receive an information item that has started spreading from a
single individual. In Figure \ref{fig:giant_size_Power}, the
curves were obtained analytically via Theorem
\ref{thm:main_Arbitrary} whereas the marked points stand for the
experimental results obtained with $n=2 \times 10^5$ nodes by averaging
$200$ experiments for each parameter set. We see that there is
an excellent agreement between theory and experiment. Moreover,
according to the experiments, the critical threshold for the
existence of a giant component (i.e., an information epidemic)
appears at $T_f \beta_f = T_w \beta_w= 0.78$ when $\alpha=0.1$,
$T_f \beta_f = T_w \beta_w= 0.61$ when $\alpha=0.5$, and $T_f
\beta_f = T_w \beta_w= 0.53$ when $\alpha=0.9$. These
values are in perfect agreement with the theoretically obtained
critical threshold $\sigma_{fw}^{\star}$ given by
(\ref{eq:sigma_star}).

The inset of Figure \ref{fig:giant_size_Power} shows
the average outbreak size $<s>$ versus $\beta_f T_f  =  \beta_w T_w$
under the same setting. 
To avoid the finite size effect (observed by Newman \cite{newman2002spread} as well)
near the epidemic threshold, we have increased the network size
up to $n=30 \times 10^6$ to obtain a better fit. Again, we see that
experimental results (obtained by averaging 
the quantity (\ref{eq:average_outbreak})
over 200 independent experiments) agree well with the
analytical results of Theorem \ref{thm:main_average_size}.

\section{Online Social Networks with $o(n)$ nodes}
\label{sec:small_social_network}

Until now, we have assumed that apart from the physical network
$\mathbb{W}$ on $n$ nodes, information can spread over an
online social network which has $\Omega(n)$ members.
However, one may also wonder as to what would happen if the number
of nodes in the online network is a sub-linear fraction of $n$.
For instance, consider an online social network $\mathbb{F}$ whose
vertices are selected by picking each node $1,\ldots, n$ with
probability $n^{\gamma-1}$ where $0<\gamma<1$. This would yield a
vertex set $\mathcal{N}_F$ that satisfies
\begin{equation}
|\mathcal{N}_F| \leq  n^{\gamma} (1+\epsilon) \label{eq:N_F_bound}
\end{equation}
with high probability for any $\epsilon>0$. We now show that, asymptotically, social
networks with $n^{\gamma}$ nodes have almost no effect in
spreading information. We start by establishing an upper bound
on the size of the giant component in $\mathbb{H}=\mathbb{W} \cup
\mathbb{F}$.

\begin{prop}
{\sl Let $\mathbb{W}$ be a graph on vertices $1,\ldots, n$, and
$\mathbb{F}$ be a graph on the vertex set $\mathcal{N}_F \subset
\{1,\ldots, n\}$. With $\mathbb{H} = \mathbb{W} \cup \mathbb{F}$,
we have
\begin{equation}
C_1(\mathbb{H}) \leq C_1(\mathbb{W}) + C_2(\mathbb{W})
(|\mathcal{N}_F|-1), \label{eq:components}
\end{equation}
where $C_1(\mathbb{W})$ and $C_2(\mathbb{W})$ are sizes of the
first and second largest components of $\mathbb{W}$,
respectively.} \label{prop:bound_on_giant_size}
\end{prop}

\proof It is clear that $C_1(\mathbb{H})$ will take its largest
value when $\mathbb{F}$ is a fully connected graph; i.e., a graph
with edges between every pair of vertices. In that case the largest
component of $\mathbb{H}$ can be obtained by taking a union of the
largest components of $\mathbb{W}$ that can be reached from the
nodes in $\mathcal{N}_F$. With $C_1^{i}(\mathbb{W})$
denoting set of nodes in the largest component (of $\mathbb{W}$)
that can be reached from node $i$, we have
\begin{equation}
C_1(\mathbb{H}) =\left|\cup_{i \in \mathcal{N}_F}
C_1^{i}(\mathbb{W})\right| \leq C_1(\mathbb{W}) + C_2(\mathbb{W})
+\ldots C_{|\mathcal{N}_F|}(\mathbb{W}) \label{eq:ineq_sharp_comp}
\end{equation}
where $C_j(\mathbb{W})$ stands for the $j$th largest component of
$\mathbb{W}$. The inequality (\ref{eq:ineq_sharp_comp}) is easy to
see once we write
\[
\left|\cup_{i \in \mathcal{N}_F} C_1^{i}(\mathbb{W})\right| =
\sum_{i=1}^{|\mathcal{N}_F|}\left|C_1^{\mathcal{N}_F(i)}(\mathbb{W})-\bigcup_{j
=1}^{i-1} C_1^{\mathcal{N}_F(j)}(\mathbb{W})\right|
\]
where $\mathcal{N}_F(i)$ is the $i$th element of $\mathcal{N}_F$.
The above quantity is a summation of the sizes of
$|\mathcal{N}_F|$ mutually disjoint components of $\mathbb{W}$. As
a result, this summation can be no larger than the sum of the
first $|\mathcal{N}_F|$ largest components of $\mathbb{W}$. The
desired conclusion (\ref{eq:components}) is now immediate as we
note that $C_2(\mathbb{W}) \geq C_j(\mathbb{W})$ for all
$j=3,\ldots,\mathcal{N}_F$.
\endproof

The next result is an easy consequence of Proposition
\ref{prop:bound_on_giant_size} and classical results
\cite{Bollobas} for ER graphs.

\begin{cor}
{\sl Let $\mathbb{W}$ be an ER graph on the vertices $1,\ldots, n$
and let $\mathbb{F}$ be a graph whose vertex set $\mathcal{N}_F$
satisfies (\ref{eq:N_F_bound}) whp. The followings hold for
 $\mathbb{H}=\mathbb{W}
\cup \mathbb{F}$:
\begin{itemize}
\item[($i)$] If $\mathbb{W}$ is in the subcritical regime (i.e.,
if $C_1(\mathbb{W}) =o(n)$), then whp we have
$C_1(\mathbb{H})=o(n)$. \item[$(ii)$] If $C_1(\mathbb{W}) =
\Theta(n)$, then we have
\[
 C_1(\mathbb{H}) = (1+o(1)) C_1(\mathbb{W}).
\]
\end{itemize}
} \label{cor:small_ER}
\end{cor}

\proof It is known \cite{Bollobas} that for an ER graph
$\mathbb{W}$, it either holds that $C_1(\mathbb{W})=O(\log n)$
(subcritical regime) or it is the case that
$C_1(\mathbb{W})=\Theta(n)$ while $C_2(\mathbb{W})=O(\log n)$
(supercritical regime). Under condition (\ref{eq:N_F_bound}),
we see from (\ref{eq:components}) that whenever $C_1(\mathbb{W})
=o(n)$ we have $C_1(\mathbb{H}) \leq c \log n \cdot n^{\gamma}$
for some $c>0$ and part $(i)$ follows immediately. Next, assume
that we have $C_1(\mathbb{W}) = \Theta(n)$. The claim $(ii)$
follows from (\ref{eq:components}) as we note that
\[
\lim_{n \to \infty } \frac{C_2(\mathbb{W}) \cdot
n^{\gamma}}{C_1(\mathbb{W})} = 0
\]
since $C_2(\mathbb{W})=O(\log n)$.
\endproof

Corollary \ref{cor:small_ER} shows that in the case where only a
sub-linear fraction of the population use online social networks,
an information item originating at a particular node can reach a positive fraction of individuals if
and only if information epidemics are already possible in the
physical information network. Moreover, we see from Corollary
\ref{cor:small_ER} that the fractional size of a possible
information epidemic in the conjoint social-physical network is
the {\em same} as that of the physical information network alone.
Combining these, we conclude that online social networks with
$n^{\gamma}$ ($0\leq \gamma < 1$) members have no effect on the
(asymptotic) fraction of individuals that can be influenced by an
information item in the conjoint social-physical network.

Along the same line as in Corollary \ref{cor:small_ER}, we 
next turn our attention to random
graphs $\mathbb{W}$ with arbitrary degree distribution
\cite{newman2001random,MolloyReed}. This time we rely on the
results by Molloy and Reed \cite[Theorem 1]{MolloyReed} who have
shown that if there exists some $\epsilon>0$ such that
\begin{equation}
\max\{d_i, i=1,\ldots, n\} \leq n^{\frac{1}{4} -\epsilon}
\label{eq:cond1}
\end{equation}
then in the supercritical regime (i.e., when $C_1(\mathbb{W}) =
\Theta(n)$) we have $C_2(\mathbb{W})= O(\log n)$. It was also
shown \cite{MolloyReed} that in the subcritical regime of the
phase transition, we have $C_1(\mathbb{W}) = O(w(n)^2 \log n)$
whenever
\begin{equation}
 \max \left\{d_i, i=1,\ldots, n \right\} \leq w(n)
 \label{eq:cond2}
\end{equation}
with $w(n) \leq n^{\frac{1}{8}-\epsilon}$ for some $\epsilon>0$.

Now, consider a graph $\mathbb{F}$ whose vertex set
$\mathcal{N}_F$ satisfies (\ref{eq:N_F_bound}) whp and let
$\mathbb{W}$ be a random graph with a given degree sequence
$\{d_i\}_{i=1}^{n}$ satisfying (\ref{eq:cond1}). In view of
(\ref{eq:components}), it is easy to see that
\[
C_1(\mathbb{H}) = (1+o(1))C_1(\mathbb{W})
\]
where $\mathbb{H}=\mathbb{W} \cup \mathbb{F}$ and this provides an
analog of Corollary \ref{cor:small_ER}(ii). It is also immediate
that in the subcritical regime, we have
\[
C_1(\mathbb{H}) = o(n)
\]
as long as (\ref{eq:cond2}) is satisfied and (\ref{eq:N_F_bound})
holds for some $\gamma \leq \frac{3}{4}$; this establishes an analog
of Corollary \ref{cor:small_ER}(i) for graphs with arbitrary
degree distributions.

The above result takes a simpler form for classes of random graphs
$\mathbb{W}$ studied in Section \ref{sec:Simu_Power}; i.e., random
graphs where the degrees follow a power-law distribution with
exponential cutoff. In particular, let the degrees of $\mathbb{W}$
be distributed according to (\ref{eq:p_k_w}). It is easy to see
that
\[
\max\{d_i, i=1,\ldots, n\} = O(\log n)
\]
with high probability so that conditions (\ref{eq:cond1}) and
(\ref{eq:cond2}) are readily satisfied. For the latter condition,
it suffices to take $w_n=O(\log n)$ so that in the subcritical
regime, we have
\[
C_1(\mathbb{W}) = O\left((\log n)^3\right).
\]
The next corollary is now an immediate consequence of Proposition
\ref{prop:bound_on_giant_size}.

\begin{cor}
{\sl Let $\mathbb{W}$ be a random graph whose degrees follow the
distribution specified in (\ref{eq:p_k_w}) and let $\mathbb{F}$ be
a graph whose vertex set $\mathcal{N}_F$ satisfies
(\ref{eq:N_F_bound}) whp. The followings hold for
 $\mathbb{H}=\mathbb{W}\cup \mathbb{F}$:
\begin{itemize}
\item[($i)$] If $C_1(\mathbb{W}) =o(n)$, then whp we
have $C_1(\mathbb{H})=o(n)$. \item[$(ii)$] If $C_1(\mathbb{W}) =
\Theta(n)$, then we have
\[
 C_1(\mathbb{H}) = (1+o(1)) C_1(\mathbb{W}).
\]
\end{itemize}
} \label{cor:small_Arbitrary}
\end{cor}

\section{Proofs of Theorem \ref{thm:main_Arbitrary} and Theorem \ref{thm:main_average_size}}
\label{sec:Proof_Arbitrary}

Consider random graphs $\mathbb{W}(n,\{p_k^w\})$ and
$\mathbb{F}(n;\alpha,\{p_k^f\})$ as in 
Section \ref{sec:Results_Arbitrary}. In order to study the diffusion of information
in the overlay network $\mathbb{H} = \mathbb{W} \cup \mathbb{F}$,
we consider a branching process which starts by giving a piece of information to 
an arbitrary node, and then recursively reveals  the largest number of nodes 
that are reached {\em and}
informed by exploring its neighbors. We remind that information propagates
from a node to each of its neighbors independently with probability $T_f$
through links in $\mathbb{F}$ (type-$1$) and with probability $T_w$ through links in
$\mathbb{W}$ (type-$2$). In the following, we utilize the standard approach on
generating functions \cite{newman2001random,newman2002spread},
and determine the condition for the existence of a giant informed component as well as
the final expected size of the information epidemics.
This approach is valid long as
the initial stages of the branching process is locally 
tree-like, which holds in this case as the clustering 
coefficient of colored degree-driven networks scales like $1/n$ as $n$ gets
large \cite{Soderberg3}.



We now solve for the survival probability of the aforementioned
branching process by using the {\em mean-field} approach based on the
generating functions \cite{newman2001random,newman2002spread}. 
Let $h_1(x)$ (resp. $h_2(x)$) denote the
generating functions for the {\em finite} number of nodes reached and informed 
by following a type-$1$ (resp. type-$2$) edge in the above branching process. 
In other words, we let $h_1(x)=\sum v_m x^m$ where $v_m$ is the \lq\lq probability 
that an arbitrary type-$1$ link leads to a {\em finite} informed component of size $m$";
$h_2(x)$ is defined analogously for type-$2$ links. Finally, we let $H(x)$ define
the generating function for the finite number of nodes that receive an information started
from an arbitrary node. 

We start by deriving the recursive relations governing $h_1(x)$ and $h_2(x)$. 
We find that 
the generating functions 
$h_1(x)$ and $h_2(x)$ satisfy the self-consistency equations
\begin{eqnarray}
h_1(x)= x \sum_{\boldsymbol{d}} \frac{d_f p_{\boldsymbol{d}}}{<d_f>}
 T_f  h_1(x)^{d_f-1} h_2(x)^{d_w}  + (1-T_f) \hspace{1mm}
\label{eq:g_1}
\\
h_2(x) = x \sum_{\boldsymbol{d}} \frac{d_w p_{\boldsymbol{d}}}{<d_w>}
 T_w  h_1(x)^{d_f} h_2(x)^{d_w-1}+
(1-T_w)
\label{eq:g_2}
\end{eqnarray}

The validity of (\ref{eq:g_1}) can be seen as follows:
The explicit factor $x$ accounts for the initial vertex that is arrived at. The factor 
$d_f p_{\boldsymbol{d}}/<d_f> $ gives \cite{newman2002spread} 
the {\em normalized} probability that an edge of 
type $1$ is attached (at the other end) to a vertex
with colored degree $\boldsymbol{d}=(d_f, d_w)$. Since the arrived node is reached 
by a type-$1$ link, it will receive the information with probability 
$T_f$. If the arrived node receives the information, 
it will be added to the component of informed nodes and
it can inform other nodes via its remaining $d_f-1$
links of type-$1$ and $d_w$ links of type-$2$. 
Since the number of nodes reached and informed by each of its
type-$1$ (resp. type-$2$) links is 
generated in turn by $h_1(x)$ 
(resp. $h_2(x)$) we obtain the 
term $h_1(x)^{d_f-1} h_2(x)^{d_w}$ 
by the powers property of generating functions
\cite{newman2001random,newman2002spread}. Averaging
over all possible colored degrees $\boldsymbol{d}$ gives the 
first term in (\ref{eq:g_1}).
The second term with the factor $x^{0}=1$ accounts for the 
possibility that the arrived node does {\em not} receive
the information and 
thus is not included in the cluster of informed nodes.
The relation (\ref{eq:g_2}) can be validated via 
similar arguments.

Using the relations (\ref{eq:g_1})-(\ref{eq:g_2}), we now find the
{\em finite} number of nodes reached and informed
by the above branching process. We have that
\begin{equation}
H(x)=x \sum_{\boldsymbol{d}} p_{\boldsymbol{d}} h_1(x)^{d_f} h_2(x)^{d_w}.
\label{eq:g_e(x)}
\end{equation}
Similar to (\ref{eq:g_1})-(\ref{eq:g_2}), the relation (\ref{eq:g_e(x)})
can be seen as follows: The factor $x$ corresponds to the 
initial node that is selected arbitrarily and given a piece of information. 
The selected node has
colored degree $\boldsymbol{d}=(d_f,d_w)$ with probability $p_{\boldsymbol{d}}$.
The number of nodes it reaches and informs via each of its $d_f$ (resp. $d_w$)
links of type $1$ (resp. type $2$)
is generated by $h_1(x)$ (resp. $h_2(x)$). 
This yields the term $h_1(x)^{d_f} h_2(x)^{d_w}$
and averaging over all possible colored degrees,
we get (\ref{eq:g_e(x)}).

We are interested in the solution of the recursive
relations (\ref{eq:g_1})-(\ref{eq:g_2}) for the case $x=1$. This case exhibits a
trivial fixed point $h_1(1) = h_2(1)=1$ which
yields $H(1)=1$ meaning that the underlying branching process is
in the subcritical regime and that {\em all} informed components have
finite size as understood from the conservation of probability.
However, the fixed point $h_1(1) = h_2(1) =1$
corresponds to the physical solution only if it is an {\em
attractor}; i.e., a stable solution to the recursion (\ref{eq:g_1})-(\ref{eq:g_2}).
The stability of this fixed point can be checked via linearization
of (\ref{eq:g_1})-(\ref{eq:g_2}) around $h_1(1) = h_2(1)=1$, which 
yields the Jacobian matrix $\boldsymbol{J}$ given by
$
\boldsymbol{J}(i,j) = \frac{\partial h_i(1)}{\partial h_j(1)}\big|_{h_1(1)=h_2(1)=1}
$
for $i,j=1,2$. This gives
\begin{equation}
\boldsymbol{J} = 
\left[
\begin{array}{cc}
\frac{T_f < (d_f^2 - d_f) >}{<d_f>} & 
\frac{T_f < d_f d_w >}{<d_f>} \\
& \\
\frac{T_w < d_w d_f >}{<d_w>} &
\frac{T_w < (d_w^2 - d_w) >}{<d_w>}
\end{array}
\right].
\label{eq:J_p}
\end{equation}

If all the eigenvalues of $\boldsymbol{J}$ are less than one in
absolute value (i.e., if the spectral radius $\sigma(\boldsymbol{J})$ 
of $\boldsymbol{J}$ satisfies $\sigma(\boldsymbol{J}) \leq 1$), 
then the solution ${h}_1(1) = h_2(1)=1$ 
is an attractor and $H(1)=1$ becomes the physical
solution, meaning that  $\mathbb{H}(n;\alpha,\{p^w_k\},T_w,\{p^f_k\},T_f)$
does not possess a giant component whp. In that case, the fraction
of nodes that receive the information tends to zero 
in the limit $n \to \infty$. On the other hand, if the
spectral radius of $\boldsymbol{J}$ is larger than one, then the
fixed point ${h}_1(1) = h_2(1)=1$ is unstable
pointing out that the asymptotic branching process is
supercritical, with a positive probability of producing infinite
trees. In that case, a nontrivial fixed point exists and becomes
the attractor of the recursions (\ref{eq:g_1})-(\ref{eq:g_2}), yielding a solution
with $h_1(1), h_2(1) < 1$. In view of (\ref{eq:g_e(x)}) this
implies $H(1) < 1$ and the corresponding probability deficit
$1-H(1)$ is attributed to the existence of a giant (infinite) 
component of informed nodes. In
fact, the quantity $1-H(1)$ is equal to the probability that a
randomly chosen vertex belongs to the giant component, which
contains asymptotically a fraction $1-H(1)$ of the vertices.

Collecting these Theorem \ref{thm:main_Arbitrary} is now within easy reach. 
First, recall that $k_f$ and $k_w$ are random variables
independently drawn from the distributions $\{p_k^f\}$ and
$\{p_k^w\}$, respectively, so that $d_f$ is a random variable that is
statistically equivalent to $k_f$ with probability $\alpha$, and
equal to zero otherwise. On the other hand,
we have $k_w \buildrel st \over = d_w$. Using
(\ref{eq:betas_new_def}) in (\ref{eq:J_p}), we now get
\begin{equation}\nonumber
\boldsymbol{J} = 
\left[
\begin{array}{cc}
T_f  \beta_f & 
T_f \lambda_w \\
T_w \alpha \lambda_f &
 T_w  \beta_w
\end{array}
\right]
\end{equation}
by the independence of $d_f$ and $d_w$. It is now a simple matter to 
see that $\sigma(\boldsymbol{J}) = \sigma_{fw}^{\star}$,
where $\sigma_{fw}^{\star}$ is as defined in (\ref{eq:sigma_star}). 
Therefore, we have established that the epidemic threshold 
is given by $\sigma_{fw}^{\star}=1$, and part (i) of Theorem \ref{thm:main_Arbitrary} follows.

Next, we set $x=1$ in the recursive relations (\ref{eq:g_1})-(\ref{eq:g_2})
and let $h_1:=h_1(1)$ and $h_2:=h_2(1)$. Using (\ref{eq:colored_dist})
and elementary algebra, we find that the stable solution of the recursions
(\ref{eq:g_1})-(\ref{eq:g_2}) is given by the smallest solution of
(\ref{eq:h_1})-(\ref{eq:h_2}) with $h_1,h_2$ in $(0,1]$. It is also easy to check from
(\ref{eq:g_e(x)}) that 
\[
H(1) = \bE{\alpha h_1^{k_f} +
1-\alpha}   \times \bE{h_2^{k_w}},
\]
and part (ii) of Theorem \ref{thm:main_Arbitrary} follows upon recalling that
the fractional size of the giant component (i.e., the number of informed nodes) 
is given by $1-H(1)$ whp. 
\endproof

We now turn to proving Theorem \ref{thm:main_average_size}. In the subcritical regime,
$H(x)$ corresponds to generating function for the distribution of 
outbreak sizes; i.e, distribution of the number of nodes that receive an information started
from an arbitrary node. Therefore, the mean outbreak size is given by the 
first derivative of $H(x)$ at the point $x=1$. Namely, we have
\begin{equation}
<s> = \frac{d H(x)}{dx} \Big|_{x=1} := H'(1).
\label{eq:<s>}
\end{equation}
Recalling that $h_1(1)=h_2(1)=1$
in the subcritical regime, 
we get from (\ref{eq:g_e(x)}) that
\begin{eqnarray}\nonumber
H'(1) &=& 1+ h_1'(1) \sum_{\boldsymbol{d}} p_{\boldsymbol{d}} d_f + 
h_2'(1)\sum_{\boldsymbol{d}} p_{\boldsymbol{d}} d_w
\\ 
&=& 1 + \alpha \lambda_f h_1'(1) + \lambda_w h_2'(1).
\label{eq:first_derivative_H}
\end{eqnarray}

The derivatives $h_1'(1)$ and $h_2'(1)$ can also be computed recursively
using the relations (\ref{eq:g_1})-(\ref{eq:g_2}). In fact, it is easy to check that
\begin{eqnarray}
h_1'(1) &=& T_f + T_f \beta_f  h_1'(1) + T_f  \lambda_w  h_2'(1)
\label{eq:h_1_derivative}
\\
h_2'(1) &=& T_w  + T_w \alpha \lambda_f h_1'(1) + T_w \beta_w  h_2'(1)
\label{eq:h_2_derivative}
\end{eqnarray}
upon using (\ref{eq:betas_new_def}) and other definitions introduced previously.
Now, setting $s_1:=h_1'(1)$ and $s_2:=h_2'(1)$, we obtain 
(\ref{eq:s_1})-(\ref{eq:s_2}) from (\ref{eq:h_1_derivative})-(\ref{eq:h_2_derivative}), and
Theorem \ref{thm:main_average_size} follows upon substituting (\ref{eq:first_derivative_H})
into (\ref{eq:<s>}).
\endproof

\section{Proof of Theorem \ref{thm:main_ER}}
\label{sec:Proof_ER}

In this section, we give a proof Theorem \ref{thm:main_ER}. First,
we summarize the technical tools that will be used.

\subsection{Inhomogeneous Random Graphs}
\label{sec:Bollobas}

Recently, Bollobas, Janson and Riordan
\cite{BollobasJansonRiordan} have developed a new theory of
inhomogeneous random graphs that would allow studying phase transition 
properties of complex networks in a rigorous fashion. The authors
in \cite{BollobasJansonRiordan} established very general results
for various properties of these models, including the critical
point of their phase transition, as well as the size of their
giant component. Here,  we summarize these tools with focus on
the results used in this paper.

At the outset, assume that a graph is defined on vertices
$\{1,\ldots,n\}$, where each vertex $i$ is assigned randomly {\em
or} deterministically a point $x_i$ in a metric space $S$. Assume
that the metric space $S$ is equipped with a Borel probability
measure $\mu$ such that for any $\mu$-continuity set $A \subseteq
S$ (see \cite{BollobasJansonRiordan})
\begin{equation}
\frac{1}{n}\cdot\sum_{i=1}^{n}\1{x_i \in A} \xrightarrow{p}
\mu(A). \label{eq:measure}
\end{equation}
A vertex space $\mathcal{V}$ is then defined as a triple $(S, \mu,
\{x_1,\ldots,x_n\})$ where $\{x_1,\ldots, x_n\}$ is a sequence of
points in $S$ satisfying (\ref{eq:measure}).

Next, let a kernel $\kappa$ on the space $(S,\mu)$ define a
symmetric, non-negative, measurable function on $S \times S$. The
random graph $G^{\mathcal{V}}(n,\kappa)$ on the vertices
$\{1,\ldots, n\}$ is then constructed by assigning an edge between
$i$ and $j$ ($i<j$) with probability $\kappa(x_i,x_j)/n$,
independently of all the other edges in the graph.

Consider random graphs $G^{\mathcal{V}}(n,\kappa)$ for which the
kernel $\kappa$ is bounded and continuous a.e. on $S \times S$. In
fact, in this study it suffices to consider only the cases where
the metric space $S$ consists of finitely many points, i.e.,
$S=\{1,\ldots,r\}$. Under these
assumptions, the kernel $\kappa$ reduces to an $r \times r$
matrix, and $G^{\mathcal{V}}(n,\kappa)$ becomes a random graph
with vertices of $r$ different types; e.g., vertices with/without
Facebook membership, etc. Two nodes (in
$G^{\mathcal{V}}(n,\kappa)$) of {\em type} $i$ and $j$ are joined by an
edge with probability $n^{-1}\kappa(i,j)$ and the condition
(\ref{eq:measure}) reduces to
\begin{equation}
\frac{n_i}{n} \xrightarrow{p} \mu_i, \quad i=1,\ldots, r,
\label{eq:measure_reduced}
\end{equation}
where $n_i$ stands for the number of nodes of type $i$ and $\mu_i$
is equal to $\mu(\{i\})$.

As usual, the phase transition properties of
$G^{\mathcal{V}}(n,\kappa)$ can be studied by exploiting the
connection between the component structure of the graph and the
survival probability of a related branching process. In
particular, consider a branching process that starts with an
arbitrary vertex and recursively reveals the largest component
reached by exploring its neighbors. For each $i=1,\ldots, r$, we
let $\rho(\kappa;i)$ denote the probability that the branching
process produces infinite trees when it starts with a node of type
$i$. The survival probability $\rho(\kappa)$ of the branching
process is then given by
\begin{equation}\label{eq:survival_prob}
\rho(\kappa)= \sum_{i=1}^{r} \rho(\kappa;i)\mu_i.
\end{equation}

In analogy with the classical results for ER graphs
\cite{Bollobas}, it is shown
\cite{BollobasJansonRiordan} that $\rho(\kappa;i)$'s
satisfy the recursive equations
\begin{equation}\label{eq:fixed_point}
\rho(\kappa;i)=1-\exp\left\{-\sum_{j=1}^r \kappa(i,j) \mu_j\cdot
\rho(\kappa;j)\right\}, \:\: i=1,\ldots, r.
\end{equation}
The value of $\rho(\kappa)$ can be computed via
(\ref{eq:survival_prob}) by characterizing the stable fixed point
of (\ref{eq:fixed_point}) reached from the starting point
$\rho(\kappa;1)=\cdots=\rho(\kappa;r)=0$. It is a simple matter to
check that, with $\boldsymbol{M}$ denoting an $r \times r$ matrix
given by $\boldsymbol{M}(i,j)= \kappa(i,j) \cdot \mu_j$, the
iterated map (\ref{eq:fixed_point}) has a non-trivial solution
(i.e., a solution other than
$\rho(\kappa;1)=\cdots=\rho(\kappa;r)=0$) {\em iff}
\begin{equation}
\sigma(\boldsymbol{M}):=\max\{|\lambda_i|:\lambda_i~\mbox{is~an~eigenvalue~of~$\boldsymbol{M}$}\}>1.
\label{eq:sigma>1}
\end{equation}
Thus, we see
that if the spectral radius of $\boldsymbol{M}$ is less than or
equal to one, the branching process is subcritical with
$\rho(\kappa)=0$ and the graph $G^{\mathcal{V}}(n,\kappa)$ has no
giant component; i.e., we have that
$C_1(G^{\mathcal{V}}(n,\kappa))=o(n)$ whp.

On the other hand, if $\sigma(\boldsymbol{M})>1$, then the
branching process is supercritical and there is a non-trivial
solution $\rho(\kappa;i)>0, i=1,\ldots, r$ that corresponds to a
stable fixed point of (\ref{eq:fixed_point}). In that case,
$\rho(\kappa)>0$ corresponds to the probability that an arbitrary
node belongs to the giant component, which asymptotically contains
a fraction $\rho(\kappa)$ of the vertices. In other words, if
$\sigma(\boldsymbol{M})>1$, we have that
$C_1(G^{\mathcal{V}}(n,\kappa))=\Omega(n)$ whp, and $\frac{1}{n}
C_1(G^{\mathcal{V}}(n,\kappa)) \xrightarrow{p} \rho(\kappa)$.

Bollobas et al. \cite[Theorem 3.12]{BollobasJansonRiordan} have
shown that the bound $C_1(G^{\mathcal{V}}(n,\kappa))=o(n)$ in the
subcritical case can be improved under some additional conditions:
They established that whenever $\sup_{i,j}\kappa(i,j) < \infty$
and $\sigma(\boldsymbol{M})\leq 1$, then we have
$C_1(G^{\mathcal{V}}(n,\kappa))=O(\log n)$ whp as in the case of
ER graphs. They have also shown that if either
$\sup_{i,j}\kappa(i,j) < \infty$ or $\inf_{i,j}\kappa(i,j) > 0$,
then in the supercritical regime (i.e., when
$\sigma(\boldsymbol{M})
> 1$) the second largest component satisfies
$C_2(G^{\mathcal{V}}(n,\kappa))=O( \log n)$ whp.

\subsection{A Proof of Theorem \ref{thm:main_ER}}

We start by studying the information spread over the network
$\mathbb{H}$ when information transmissibilities $T_w$ and $T_f$
are both equal to $1$. Clearly, this corresponds to studying the
phase transition in
$\mathbb{H}=\mathbb{H}(n;\alpha,\lambda_w,\lambda_f)$, and we will
do so by using the techniques summarized in the previous section.
Let $S=\{1,2\}$ stand for the space of {\em vertex types}, where
vertices with Facebook membership are referred to as type $1$
while vertices without Facebook membership are said to be of type
$2$; notice that this is different than the case in the proof of Theorem
\ref{thm:main_Arbitrary} where we distinguish between different {\em link} types.
In other words, we let
\[
x_i = \left \{
\begin{array}{c}
  1 \quad \textrm{if $i \in \mathcal{N}_F$}\\
  2 \quad \textrm{if $i \not \in \mathcal{N}_F$} \\
\end{array}
\right.
\]
for each $i=1,\ldots, n$. Assume that the metric space $S$ is
equipped with a probability measure $\mu$ that satisfies the
condition (\ref{eq:measure_reduced}); i.e.,
$\mu(\{1\}):=\mu_1=\alpha$ and $\mu(\{2\}):=\mu_2=1-\alpha$.
Finally, we compute the appropriate kernel $\kappa$ such that,
for each $i,j =\{1, 2\}$, $\kappa(i,j)/n$ gives the probability that two vertices of type
$i$ and $j$ are connected. Clearly, we have
$\kappa(1,1) = n\left(1 -
\left(1-\frac{\lambda_w}{n}\right)\left(1-\frac{\lambda_f}{\alpha
n}\right)\right) = \lambda_w+\frac{\lambda_f}{\alpha} - \frac{\lambda_w
\lambda_f}{ \alpha n}$,
whereas
$
\kappa(1,2)=\kappa(2,1)=\kappa(2,2) = \lambda_w.
$

We are now in a position to derive the critical point of the phase
transition in $\mathbb{H}(n;\alpha,\lambda_w,\lambda_f)$ as well
as the giant component size
$C_1(\mathbb{H}(n;\alpha,\lambda_w,\lambda_f))$. First, we compute
the matrix $\boldsymbol{M}(i,j)=\kappa(i,j)\mu_j$ and get
\[
\boldsymbol{M} =\left[\begin{array}{cc}
\alpha \lambda_w + \lambda_f - \frac{\lambda_w \lambda_f}{n}   & (1-\alpha) \lambda_w  \\
\alpha \lambda_w  & (1-\alpha) \lambda_w \\
\end{array}\right].
\]
It is clear that the term $\frac{\lambda_w \lambda_f}{n}$
has no effect on the results as we eventually let $n$ go to
infinity. It is now a simple matter to check that the spectral radius of
$\boldsymbol{M}$ is given by
\begin{eqnarray}
\sigma(\boldsymbol{M})= \frac{1}{2}\left(\lambda_f+\lambda_w +\sqrt{\left(\lambda_f+\lambda_w
\right)^2-4(1-\alpha)\lambda_f\lambda_w}\right)
\nonumber
\end{eqnarray}

This leads to the conclusion that the random graph
$\mathbb{H}(n;\alpha,\lambda_w,\lambda_f)$ has a giant component
if and only if
\begin{equation}
\frac{1}{2}\left(\lambda_f+\lambda_w +\sqrt{\left(\lambda_f+\lambda_w
\right)^2-4(1-\alpha)\lambda_f\lambda_w}\right) > 1
\label{eq:lambda_fw_star}
\end{equation}
as we recall (\ref{eq:sigma>1}). If condition
(\ref{eq:lambda_fw_star}) is not satisfied, then we have
$C_1(\mathbb{H}(n;\alpha,\lambda_w,\lambda_f))=O(\log n)$ as we
note that $\sup_{i,j}\kappa(i,j)< \infty$. From \cite[Theorem
3.12]{BollobasJansonRiordan}, we also get that
$C_2(\mathbb{H}(n;\alpha,\lambda_w,\lambda_f))=O(\log n)$ whenever
(\ref{eq:lambda_fw_star}) is satisfied.

Next, we compute the size of the giant component whenever it
exists. Let $\rho(\kappa;1)=\rho_1$ and $\rho(\kappa;2)=\rho_2$.
In view of (\ref{eq:survival_prob}) and the arguments presented previously, 
the asymptotic fraction of nodes in the
giant component is given by
\begin{equation}
\rho(\kappa) = \alpha \rho_1 + (1-\alpha) \rho_2,
\label{eq:ro_kappa}
\end{equation}
 where $\rho_1$
and $\rho_2$ constitute a stable simultaneous solution of the transcendental
equations
\begin{equation}\begin{array}{l}
  \rho_1 = 1-\exp\left\{- \rho_1(\alpha \lambda_w+\lambda_f)-\rho_2 ( 1 - \alpha ) \lambda_w \right\} \\
 \rho_2 = 1-\exp\left\{ -\rho_1 \alpha\lambda_w -  \rho_2 ( 1 - \alpha ) \lambda_w\right\} \\
\end{array}
\label{eq:ro_1_2_ER}
\end{equation}

So far, we have established the epidemic threshold and the size of
the information epidemic when $T_w=T_f=1$. In the more general
case where there is no constraint on the transmissibilities, we
see that the online social network $\mathbb{F}$ becomes an ER
graph with average degree $T_f \lambda_f$, whereas the physical
network $\mathbb{W}$ becomes an ER graph with average degree $T_w
\lambda_w$. Therefore, the critical threshold and the size of the
information epidemic can be found by substituting $T_f \lambda_f$ for
$\lambda_f$ and $T_w \lambda_w$ for $\lambda_w$ in the
relations (\ref{eq:lambda_fw_star}), (\ref{eq:ro_kappa}) 
and (\ref{eq:ro_1_2_ER}). This establishes
Theorem \ref{thm:main_ER}.
\endproof

\section{Conclusion}
In this paper, we characterized the critical threshold and
the asymptotic size of information epidemics in an overlaying
social-physical network. To capture the spread of
information, we considered a physical
information network that characterizes the face-to-face interactions of human beings, and 
some overlaying online social networks (e.g., Facebook,
Twitter, etc.) that are defined on a subset of the population.
Assuming that information is transmitted between individuals
according to the SIR model, we showed that the critical point and
the size of information epidemics on this overlaying
social-physical network can be precisely determined.

To the best of our knowledge, this study marks the first work on
the phase transition properties of conjoint networks where the
vertex sets are neither identical (as in
\cite{ChoGohKim,KurantThiran}) nor disjoint (as in
\cite{LeichtDSouza}).  We believe that our findings here shed
light on the further studies on information (and influence) propagation across
social-physical networks.

%
%
%


\bibliographystyle{ieeetr}
\bibliography{references}{}

\begin{thebibliography}{10}

\bibitem{CPS}
{CPS Steering Group}, ``Cyber-physical systems executive summary,'' 2008.

\bibitem{Buldyrev}
S.~V. Buldyrev, R.~Parshani, G.~Paul, H.~E. Stanley, and S.~Havlin,
  ``Catastrophic cascade of failures in interdependent networks,'' {\em
  Nature}, vol.~464, pp.~1025--1028, 2010.

\bibitem{ChoGohKim}
W.~Cho, K.~I. Goh, and I.~M. Kim, ``Correlated couplings and robustness of
  coupled networks,'' {\em arXiv:1010.4971v1 [physics.data-an]}, 2010.

\bibitem{CohenHavlin}
R.~Cohen and S.~Havlin, {\em Complex Networks: Structure, Robustness and
  Function}.
\newblock United Kingdom: Cambridge University Press, 2010.

\bibitem{Vespignani}
A.~Vespignani, ``Complex networks: The fragility of interdependency,'' {\em
  Nature}, vol.~464, pp.~984--985, 2010.

\bibitem{YaganQianZhangCochranLong}
O.~Ya\u{g}an, D.~Qian, J.~Zhang, and D.~Cochran, ``\mbox{Optimal}
  \mbox{Allocation} of \mbox{Interconnecting} \mbox{Links} in
  \mbox{Cyber-Physical} \mbox{Systems}: \mbox{Interdependence},
  \mbox{Cascading} \mbox{Failures} and \mbox{Robustness},'' {\em IEEE
  Transactions on Parallel and Distributed Systems}, vol.~23, no.~9,
  pp.~1708--1720, 2012.
\newblock 

\bibitem{WebSite}
R.~L. Hotz, ``Decoding our chatter,'' {\em Wall Street Journal}, 1 October
  2011.

\bibitem{LermanGhosh}
K.~Lerman and R.~Ghosh, ``Information contagion: An empirical study of the
  spread of news on digg and twitter social networks,'' in {\em Proceedings of
  4th International Conference on Weblogs and Social Media (ICWSM)}, 2010.

\bibitem{KimYoneki}
H.~Kim and E.~Yoneki, ``Influential neighbours selection for information
  diffusion in online social networks,'' in {\em Proceedings of IEEE
  International Conference on Computer Communication Networks (ICCCN)},
  (Munich, Germany), July 2012.

\bibitem{LinBarabasi}
D.~Wang, Z.~Wen, H.~Tong, C.-Y. Lin, C.~Song, and A.-L. Barab\'{a}si,
  ``Information spreading in context,'' in {\em Proceedings of the 20th
  international conference on World wide web}, WWW '11, (New York, NY, USA),
  pp.~735--744, 2011.

\bibitem{BakshyRosennMarlowAdamic}
E.~Bakshy, I.~Rosenn, C.~Marlow, and L.~Adamic, ``The role of social networks
  in information diffusion,'' in {\em Proceedings of ACM WWW}, (Lyon, France),
  April 2012.

\bibitem{Liben-NowellKleinberg}
D.~Liben-Nowell and J.~Kleinberg, ``{Tracing information flow on a global scale
  using Internet chain-letter data},'' {\em Proceedings of the National Academy
  of Sciences}, vol.~105, pp.~4633--4638, Mar. 2008.

\bibitem{LeskovecMcGlohonFaloutsos}
J.~Leskovec, M.~McGlohon, C.~Faloutsos, N.~S. Glance, and M.~Hurst, ``Patterns
  of cascading behavior in large blog graphs,'' in {\em Proceedings of the
  Seventh SIAM International Conference on Data Mining, April 26-28, 2007,
  Minneapolis, Minnesota, USA}, 2007.

\bibitem{AndersonMay}
R.~M. Anderson and R.~M. May, {\em Infectious Diseases of Humans}.
\newblock Oxford (UK): Oxford University Press, 1991.

\bibitem{Bailey}
N.~T.~J. Bailey, {\em The Mathematical Theory of Infectious Diseases and its
  Applications}.
\newblock New York (NY): Hafner Press, 1975.

\bibitem{Hethcote}
H.~W. Hethcote, ``Mathematics of infectious diseases,'' {\em SIAM Review},
  vol.~42, no.~4, pp.~599--653, 2000.

\bibitem{KupermanAbramson}
M.~Kuperman and G.~Abramson, ``Small world effect in an epidemiological
  model,'' {\em Phys. Rev. Lett.}, vol.~86, no.~13, pp.~2909--2912, 2001.

\bibitem{MooreNewman}
C.~Moore and M.~E.~J. Newman, ``Epidemics and percolation in small-world
  networks,'' {\em Phys. Rev. E}, vol.~61, no.~5, pp.~5678--5682, 2000.

\bibitem{Pastor-SatorrasVespignani}
R.~Pastor-Satorras and A.~Vespignani, ``Epidemic spreading in scale-free
  networks,'' {\em Phys. Rev. Lett.}, vol.~86, pp.~3200--3203, 2001.

\bibitem{newman2002spread}
M.~E.~J. Newman, ``Spread of epidemic disease on networks,'' {\em Phys. Rev.
  E}, vol.~66, no.~1, 2002.

\bibitem{Blogspace}
D.~Gruhl, R.~Guha, D.~Liben-Nowell, and A.~Tomkins, ``Information diffusion
  through blogspace,'' in {\em Proceedings of the 13th international conference
  on World Wide Web}, WWW '04, (New York, NY, USA), pp.~491--501, ACM, 2004.

\bibitem{Worms}
X.~Sun, Y.-H. Liu, B.~Li, J.~Li, J.-W. Han, and X.-J. Liu, ``Mathematical model
  for spreading dynamics of social network worms,'' {\em Journal of Statistical
  Mechanics: Theory and Experiment}, vol.~2012, no.~04, p.~P04009, 2012.

\bibitem{P2P}
K.~Leibnitz, T.~Hossfeld, N.~Wakamiya, and M.~Murata, ``On pollution in
  edonkey-like peer-to-peer file-sharing networks,'' {\em Measuring, Modelling
  and Evaluation of Computer and Communication Systems (MMB), 2006 13th GI/ITG
  Conference}, pp.~1 --18, march 2006.

\bibitem{OstilliYonekiLeungMendes}
M.~Ostilli, E.~Yoneki, I.~X.~Y. Leung, J.~F.~F. Mendes, P.~Li{\`o}, and
  J.~Crowcroft, ``Statistical mechanics of rumour spreading in network
  communities,'' {\em Procedia CS}, vol.~1, no.~1, pp.~2331--2339, 2010.

\bibitem{KurantThiran}
M.~Kurant and P.~Thiran, ``Layered complex networks,'' {\em Phys. Rev. Lett.},
  vol.~96, no.~13, 2006.

\bibitem{MarceauNoelAllard}
V.~Marceau, P.-A. No{\"e}l, L.~H\'{e}bert-Dufresne, A.~Allard, and L.~J.
  Dub\'{e}, ``Modeling the dynamical interaction between epidemics on overlay
  networks,'' {\em Physical Review E}, vol.~84, no.~2, 2011.

\bibitem{LeichtDSouza}
E.~A. Leicht and R.~M. D'Souza, ``Percolation on interacting networks,'' {\em
  arXiv:0907.0894v1 [cond-mat.dis-nn]}, 2009.

\bibitem{MendiolaSerranoBoguna}
A.~Saumell-Mendiola, M.~{\'A}. Serrano, and M.~Bogu{\~n}{\'a}, ``Epidemic
  spreading on interconnected networks,'' {\em arXiv:1202.4087}, 2012.

\bibitem{Epstein}
J.~M. Epstein, ``Why model?,'' {\em Journal of Artificial Societies and Social
  Simulation}, vol.~11, no.~4, p.~12, 2008.

\bibitem{KleinbergInBook}
J.~Kleinberg, {\em {Cascading Behavior in Networks: Algorithmic and Economic
  Issues}}, ch.~24.
\newblock Cambridge University Press, 2007.

\bibitem{newman2001random}
M.~E.~J. Newman, S.~H. Strogatz, and D.~J. Watts, ``Random graphs with
  arbitrary degree distributions and their applications,'' {\em Phys. Rev. E},
  vol.~64, no.~2, 2001.

\bibitem{BroadbentHammersley}
S.~R. Broadbent and J.~M. Hammersley, ``Percolation processes,'' {\em
  Mathematical Proceedings of the Cambridge Philosophical Society}, vol.~53,
  pp.~629--641, 1957.

\bibitem{Bollobas}
B.~Bollob\'{a}s, {\em Random Graphs}.
\newblock Cambridge (UK): Cambridge Studies in Advanced Mathematics, Cambridge
  University Press, 2001.

\bibitem{ClausetShaliziNewman}
A.~Clauset, C.~R. Shalizi, and M.~E.~J. Newman, ``Power-law distributions in
  empirical data,'' {\em SIAM Rev.}, vol.~51, pp.~661--703, Nov. 2009.

\bibitem{MolloyReed}
M.~Molloy and B.~Reed, ``A critical point for random graphs with a given degree
  sequence,'' {\em Random Structures and Algorithms}, vol.~6, pp.~161--179,
  1995.

\bibitem{Soderberg2}
B.~S\"{o}derberg, ``Random graphs with hidden color,'' {\em Phys. Rev. E},
  vol.~68, no.~015102(R), 2003.

\bibitem{Grassberger}
P.~Grassberger, ``On the critical behavior of the general epidemic process and
  dynamical percolation,'' {\em Mathematical Biosciences}, vol.~63, no.~2,
  pp.~157--172, 1983.

\bibitem{BollobasJansonRiordan}
B.~Bollob\'{a}s, S.~Janson, and O.~Riordan, ``The phase transition in
  inhomogeneous random graphs,'' {\em Random Structures and Algorithms},
  vol.~33, no.~1, pp.~3--122, 2007.

\bibitem{Soderberg3}
B.~S\"{o}derberg, ``Properties of random graphs with hidden color,'' {\em Phys.
  Rev. E}, vol.~68, no.~2, pp.~026107--, 2003.

\end{thebibliography}

\appendix[Information Diffusion with Multiple Online Social Networks]
 \label{sec:multiple}

 \setcounter{equation}{0}
\renewcommand{\theequation}{\thesection.\arabic{equation}}
\renewcommand{\theprop}{\thesection.\arabic{prop}}
\renewcommand{\thecor}{\thesection.\arabic{cor}}

So far, we have assumed that information diffuses amongst human
beings via only a physical information network $\mathbb{W}$ and an
online social network $\mathbb{F}$. To be more general, one can
extend this model to the case where there are multiple online
social networks. For instance assume that there is an additional
online social network, say Twitter, denoted by
$\mathbb{T}(n;\alpha_t)$ whose members are selected by picking
each node $1,\ldots, n$ independently with probability $\alpha_t$.
In other words, with $\mathcal{N}_{T}$ denoting the set vertices
of $\mathbb{T}$, we have
\[
\bP{i \in \mathcal{N}_{T}} = \alpha_t, \quad i=1,\ldots, n.
\]
To be consistent with this notation, we assume that the members of
the online social network $\mathbb{F}$ (i.e., Facebook) are
selected by picking each node $1,\ldots, n$ independently with
probability $\alpha_f$.

The overlaying social-physical network now consists of a network
formed by conjoining $\mathbb{W}$, $\mathbb{F}$ and $\mathbb{T}$;
i.e., we have $\mathbb{H} = \mathbb{W} \cup \mathbb{F} \cup
\mathbb{T}$.
%
To demonstrate the applicability of the techniques for this
general set-up, we now consider a simple example where
$\mathbb{W}$, $\mathbb{F}$ and $\mathbb{T}$ are all ER graphs with
edge probabilities given by $\frac{\lambda_w}{n}$,
$\frac{\lambda_f}{\alpha_f n}$, and $\frac{\lambda_t}{\alpha_t
n}$, respectively. This yields an asymptotic mean degree of
$\lambda_w$, $\lambda_f$ and $\lambda_t$, in the networks
$\mathbb{W}$, $\mathbb{F}$ and $\mathbb{T}$, respectively. For the
time being, assume that the transmissibilities $T_w$, $T_f$ and
$T_t$ are all equal to one.

Now, recall the concept of inhomogeneous random graphs presented
in Section \ref{sec:Bollobas}. Let $S=\{1,2,3,4\}$ stand for the
space of vertex types, where vertices with Facebook {\em and}
Twitter membership are referred to as type $1$, vertices with
Facebook membership but {\em without} Twitter membership are
referred to as type $2$, vertices with Twitter membership but
without Facebook membership are referred to as type $3$, and
finally vertices with neither Facebook nor Twitter membership are
said to be of type $4$. That is, we set
\[
x_i = \left \{
\begin{array}{c}
  1 \quad \textrm{if $i \in \mathcal{N}_F$ \:and\: $i \in \mathcal{N}_T$}\\
  2 \quad \textrm{if $i \in \mathcal{N}_F$ \:and\: $i \not \in \mathcal{N}_T$} \\
  3 \quad \textrm{if $i \not \in \mathcal{N}_F$ \:and\: $i \in \mathcal{N}_T$}\\
  4 \quad \textrm{if $i \not \in \mathcal{N}_F$ \:and\: $i \not \in \mathcal{N}_T$} \\
\end{array}
\right.
\]
for each $i=1,\ldots, n$. Assume that the metric space $S$ is
equipped with a probability measure $\mu$ that satisfies 
condition (\ref{eq:measure_reduced}); i.e., $\mu_1=\alpha_f
\alpha_t$, $\mu_2=\alpha_f (1-\alpha_t)$, $\mu_3=(1-\alpha_f)
\alpha_t$, and $\mu_4=(1-\alpha_f) (1-\alpha_t)$. The next step is
to compute the appropriate kernel $\kappa$ such that, for each
$i, j=\{1, 2, 3, 4\}$,
$\kappa(i,j)/n$ gives the probability that two vertices of type
$i$ and $j$ are connected. For $n$ large, it is not difficult to
see that we have
\begin{eqnarray}\nonumber
\boldsymbol{\kappa}= \left[ \begin{array}{lccr}
  \lambda_w +\frac{\lambda_f}{\alpha_f} +\frac{\lambda_t}{\alpha_t} & \lambda_w +\frac{\lambda_f}{\alpha_f} & \lambda_w +\frac{\lambda_t}{\alpha_t} & \lambda_w
 \\
  \lambda_w +\frac{\lambda_f}{\alpha_f} & \lambda_w +\frac{\lambda_f}{\alpha_f} & \lambda_w & \lambda_w \\
 \lambda_w +\frac{\lambda_t}{\alpha_t} & \lambda_w & \lambda_w +\frac{\lambda_t}{\alpha_t} & \lambda_w \\
  \lambda_w & \lambda_w & \lambda_w & \lambda_w \\
\end{array}\right]
\end{eqnarray}

The matrix $\boldsymbol{M}(i,j)=\kappa(i,j)\mu_j$  is now given by
\begin{eqnarray}
\boldsymbol{M} &=& \left[ \begin{array}{ll}
  \lambda_w\alpha_f \alpha_t +  \lambda_f \alpha_t +
  \lambda_t \alpha_f & \lambda_w\alpha_f \alpha_t +
  \lambda_f \alpha_t \\
  \lambda_w\alpha_f \alpha_t +  \lambda_f \alpha_t  & \lambda_w\alpha_f \alpha_t +
  \lambda_f \alpha_t \\
  \lambda_w\alpha_f \alpha_t +  \lambda_t \alpha_f  & \lambda_w\alpha_f \alpha_t  \\
  \lambda_w\alpha_f \alpha_t & \lambda_w\alpha_f \alpha_t \\
\end{array}  \right.
\nonumber \\
 & & \quad \left. \begin{array}{ll}
\lambda_w\alpha_f \alpha_t +  \lambda_t \alpha_f  & \lambda_w\alpha_f \alpha_t \\
\lambda_w\alpha_f \alpha_t & \lambda_w\alpha_f \alpha_t  \\
\lambda_w\alpha_f \alpha_t +
  \lambda_f \alpha_t & \lambda_w\alpha_f \alpha_t  \\
 \lambda_w\alpha_f \alpha_t & \lambda_w\alpha_f \alpha_t  \\
\end{array}
 \right]
 \label{eq:M_multiple}
\end{eqnarray}
and the critical point of the phase transition as well as the
giant component size of
$\mathbb{H}(n;\alpha_f,\alpha_t,\lambda_w,\lambda_f,\lambda_t)$
can now be obtained by using the arguments of Section
\ref{sec:Bollobas}. An item of information originating from
a single node in $\mathbb{H}=\mathbb{W} \cup \mathbb{F} \cup
\mathbb{T}$ can reach a positive fraction of the individuals
only if the spectral radius of
$\boldsymbol{M}$ is greater than unity. If it is the case
that $\sigma(\boldsymbol{M}) \leq 1$, then there is no information
epidemic and all information outbreaks have size $O(\log n)$.

The fractional size of the giant component (i.e., information
epidemic) can also be found. Recalling (\ref{eq:survival_prob})
and (\ref{eq:fixed_point}), we see that
\begin{eqnarray}
\lefteqn{\frac{1}{n}
C_1\left(\mathbb{H}(n;\alpha_f,\alpha_t,\lambda_w,\lambda_f,\lambda_t)\right)}
&& \nonumber \\
&\xrightarrow{p}& \alpha_f \alpha_t \rho_1 +\alpha_f (1-\alpha_t)
\rho_2+
(1-\alpha_f) \alpha_t \rho_3 \nonumber \\
&& ~ + (1-\alpha_f) (1-\alpha_t)\rho_4, \label{eq:C_1_multiple}
\end{eqnarray}
where $0 \leq \rho_1,\rho_2,\rho_3,\rho_4 \leq 1$ are given by the
largest solution to the recursive relations
\begin{eqnarray}
\rho_i= 1-
\exp\left\{-\sum_{j=1}^{4}\boldsymbol{M}(i,j)\rho_j\right\}, \quad
i=1,2,3,4. \label{eq:rho_multiple}
\end{eqnarray}

In the case where there is no constraint on the transmissibilities
$T_w$, $T_f$ and $T_t$, the conclusions (\ref{eq:M_multiple}),
(\ref{eq:C_1_multiple}) and (\ref{eq:rho_multiple}) still apply
if we substitute $T_w \lambda_w$ for $\lambda_w$, $T_f \lambda_f$ for
$\lambda_f$ and $T_t \lambda_t$ for $\lambda_t$.

\end{document}